\documentclass[aip,jcp,
amsmath,amssymb,
reprint,
]{revtex4-1}

\usepackage{graphicx}
\usepackage{dcolumn}
\usepackage{bm}

\usepackage[utf8]{inputenc}
\usepackage[T1]{fontenc}
\usepackage{mathptmx}
\usepackage{etoolbox}
\usepackage{longtable}
\usepackage{multirow}

\makeatletter
\def\@email#1#2{%
 \endgroup
 \patchcmd{\titleblock@produce}
  {\frontmatter@RRAPformat}
  {\frontmatter@RRAPformat{\produce@RRAP{*#1\href{mailto:#2}{#2}}}\frontmatter@RRAPformat}
  {}{}
}%
\makeatother
\begin{document}


\title[Viscosity and Stokes-Einstein relation in deeply supercooled water under pressure]{Viscosity and Stokes-Einstein relation in deeply supercooled water under pressure}
\author{Alexandre Mussa}

\author{Romain Berthelard}
\thanks{A.M. and R.B. contributed equally to this work.}%

\author{Fr\'{e}d\'{e}ric Caupin}

\author{Bruno Issenmann}
\email{frederic.caupin@univ-lyon1.fr, bruno.issenmann@univ-lyon1.fr}

\affiliation{%
{Institut Lumi\`ere Mati\`ere, Universit\'e de Lyon, Universit\'e Claude Bernard Lyon 1, CNRS, Institut Universitaire de France, F-69622 Villeurbanne, France
}%
}

\date{18 September 2023}

\begin{abstract}
We report measurements of the shear viscosity $\eta$ in water up to $150\,\mathrm{MPa}$ and down to $229.5\,\mathrm{K}$. This corresponds to more than $30\,\mathrm{K}$ supercooling below the melting line. The temperature dependence is non-Arrhenius at all pressures, but its functional form at $0.1\,\mathrm{MPa}$ is qualitatively different from that at all pressures above $20\,\mathrm{MPa}$. The pressure dependence is non-monotonic, with a pressure-induced decrease of viscosity by more than 50\% at low temperature. Combining our data with literature data on the self-diffusion coefficient $D_\mathrm{s}$ of water, we check the Stokes-Einstein relation which, based on hydrodynamics, predicts constancy of $D_\mathrm{s} \eta/T$, where $T$ is the temperature. The observed temperature and pressure dependence of $D_\mathrm{s} \eta/T$ is analogous to that obtained in simulations of a realistic water model. This analogy suggests that our data are compatible with the existence of a liquid-liquid critical point at positive pressure in water.
\end{abstract}
\maketitle

\section{\label{sec:Introduction} Introduction}

Among the numerous anomalies of water, its shear viscosity $\eta$ shows an intriguing non-monotonous behavior, reaching a minimum value upon increasing pressure along an isotherm. Already observable in the stable liquid \cite{Bett1966}, this anomaly becomes more pronounced in the supercooled liquid \cite{Singh2017}, where we measured it recently down to 244.3\,K. 

In general, the shear viscosity is tightly coupled to the molecular self-diffusion coefficient $D_\mathrm{s}$ through the Stokes-Einstein relation (SER) which states that the quantity $r_\mathrm{SE}=D_\mathrm{s}\eta/T$, where $T$ is the absolute temperature, remains constant. In most glassformers, $r_\mathrm{SE}$ is indeed nearly constant down to very low temperatures, and begins increasing only below around $1.2T_g$, where $T_g$ is the glass transition temperature \cite{Chang1997}. In light\cite{Dehaoui2015} and heavy\cite{Ragueneau2022} water however, the SER is already violated at room temperature, more than twice the glass transition temperature. Currently, the microscopic explanation for this behavior is the increasing importance of translational jump motion in diffusion of the water molecules as the liquid becomes more supercooled\cite{Dubey2019}.

The effect of pressure on the violation of the SER in water has been investigated by molecular dynamics simulations. Many water models exhibit a first-order liquid-liquid transition (LLT) between two liquids differing in density and structure. The LLT would terminate at a liquid-liquid critical point (LLCP) at a pressure $P_\mathrm{c}$ in the supercooled liquid\cite{Gallo2016}. From the LLCP emanates a Widom line, i.e. a line of correlation length maxima associated to the LLT, located at temperature $T_\mathrm{W}(P)$ at pressure $P$. Based on their simulations, Kumar \textit{et al.}\cite{Kumar2007} found a connection between the violation of the SER, the LLT, and the Widom line. While violated at high temperatures at pressures below $P_\mathrm{c}$, the SER holds to lower temperatures above $P_\mathrm{c}$. Furthermore, below $P_\mathrm{c}$, the curves showing $r_\mathrm{SE}$ at various $P$ collapse on a single curve when plotted as a function of $T-T_\mathrm{W}(P)$. A limitation of Ref.~\onlinecite{Kumar2007} is the use of the structural relaxation time $\tau_\alpha$ as a proxy for $\eta$ in the SER. Later work which directly simulated $\eta$\cite{Montero2018,Dubey2019} gave a less clear-cut picture. Above $P_\mathrm{c}$, violation of SER is still observed, but with no pressure evolution in the considered temperature range; below $P_\mathrm{c}$, the collapse of $r_\mathrm{SE}$ plotted as a function of $T-T_\mathrm{W}(P)$ is not perfect.

These observations call for experimental data to investigate the SER in supercooled water under pressure. While experimental $D_\mathrm{s}$ values are available in a broad temperature and pressure range including the supercooled region (up to 400\,MPa and 20 to 60\,K supercooling depending on the pressure), data on $\eta$ are scarce. Indeed, until recently, only two sets\cite{Hallet1963,Osipov1977} of viscosity measurements in supercooled water were available, and only at atmospheric pressure. In the past years, we developed two experimental setups aimed at viscosity measurement in supercooled water. The first one, based on Brownian motion of colloidal spheres, gave reliable values for $\eta$ under atmospheric pressure and down to $239.3\,\mathrm{K}$ for H$_2$O\cite{Dehaoui2015} and $243.7\,\mathrm{K}$ for D$_2$O\cite{Ragueneau2022}, confirming a bias in one of the previous data set~\cite{Osipov1977}, as was already suspected \cite{Cho1999,Dehaoui2015}. The second setup, involving Poiseuille flow in a high pressure capillary, yielded the first measurements of shear viscosity in supercooled H$_2$O under pressure, up to $300\,\mathrm{MPa}$\cite{Singh2017}. However, the degree of supercooling was limited to around 20\,K supercooling, due to heterogeneous ice nucleation in the moderately large, flowing water sample.

Here we report $\eta$ values in H$_2$O under pressure at even lower temperature, obtained from Brownian motion of colloidal spheres. The experimental details and procedures are given in Section~\ref{sec:MatMet}, and the results presented in Section~\ref{sec:results}. They enable a detailed study of the experimental SER in supercooled water under pressure (Section~\ref{sec:SER}). The findings are discussed in Section~\ref{sec:disc} in the light of a comparison with molecular dynamics simulations.

\begin{figure*}
    \centering
    \includegraphics[width=16cm]{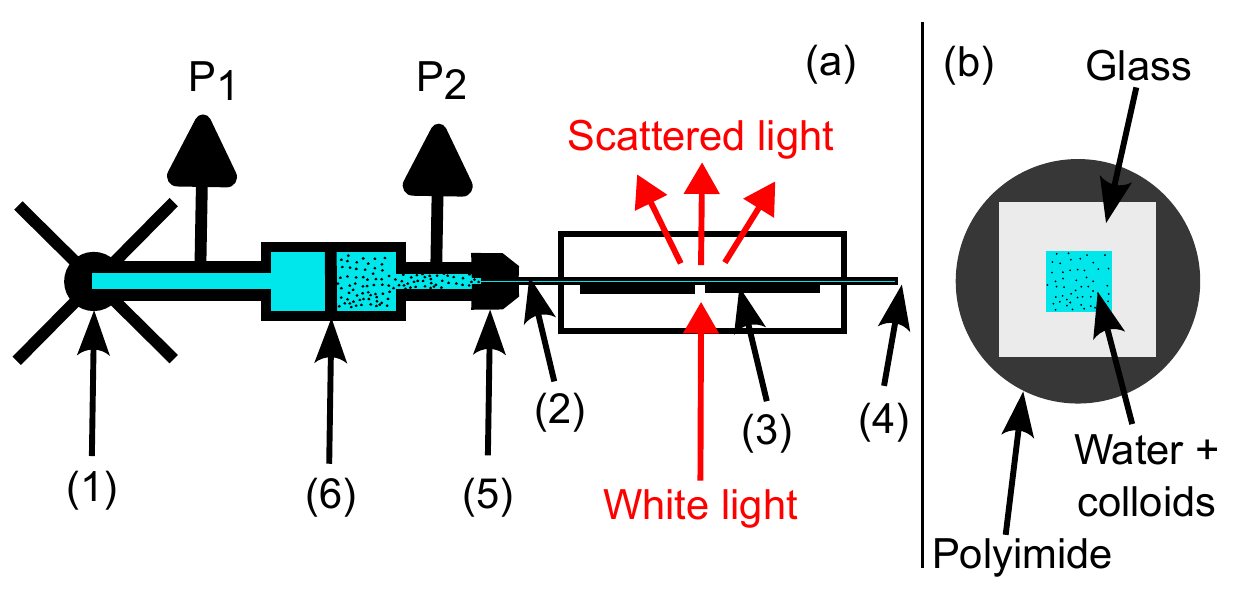}
    \caption{Experimental setup. (a) High pressure system: A screw pump (1) controls the sample pressure up to $200\,\mathrm{MPa}$. The sample is a suspension of polystyrene colloids in ultrapure water held in a square capillary (2). Its temperature is controlled by a Linkam stage whose silver plate (3) is cooled by nitrogen gas. The capillary is flame-sealed at one end (4) and connected at the other end (5) to the high pressure system. A separating piston (6) prevents colloids to invade the high pressure setup. The pressures $P_1$ and $P_2$ are measured on both sides of the piston. A microscope in transmission mode collects the white light scattered by the sample, and a time-correlation analysis yields the diffusion coefficient of the colloids. (b) Cross section of the square fused-silica capillary (see text for details).}
    \label{fig:ExperimentalSetup}
\end{figure*}

\section{Materials and methods\label{sec:MatMet}}

\subsection{Experimental setup}

As described previously\cite{Dehaoui2015, Ragueneau2022}, we use a colloidal suspension of polystyrene particles (Duke Scientific 3000 Series), rinsed and diluted 100 times in supercooled ultrapure water to reach a volume fraction of $10^{-4}$. A microscope (Zeiss Axioscope) with a $50\times$ objective (Mitutoyo, Plan Apo, A.N. 0.55) and a CCD camera (Prosilica GX1050, Allied Vision Technologies) records 500 image movies, the duration of which is varied depending on the viscosity of water (from 5 to $87\,\mathrm{s}$). A Matlab code uses the decorrelation of images due to the Brownian motion of the colloids to deduce their diffusion coefficient $D$ following Refs.~\onlinecite{Cerbino2008, Giavazzi2009}. We deduce the viscosity $\eta$ of water using the Stokes-Einstein relation $\eta=k_\mathrm{B}T/(6\pi D a)$, where $k_\mathrm{B}$ is the Boltzmann constant, $T$ the temperature, $a$ the radius of the colloids. In the measurement at $100$ and $150$ MPa the diameter of the colloids was $2a = 203\pm 5\,\mathrm{nm}$ while at lower pressures, it was $2a = 347\pm 6\,\mathrm{nm}$.

The sample container was modified for high pressure measurement (Fig.~\ref{fig:ExperimentalSetup}). The colloidal suspension was held in a thick-wall, fused silica tubing (Polymicro WWP050375). The cross-section of the capillary is a square with inner dimension $(50 \pm 5)\,\mathrm{\mu m}$, and its wall thickness is $(160 \pm 10)\,\mathrm{\mu m}$. It is protected with a cylindrical polyimide coating, that was burnt before experiments along 5 to 10\,mm around the observation area. A $(225\pm 5)\,\mathrm{mm}$ long section of the capillary is cut and filled with the solution by capillarity. One end is flame sealed, and the other connected to a high pressure circuit through a fitting similar to the one described in Ref.~\onlinecite{Patel2004}.

The capillary is then inserted through a Linkam CAP500 thermal stage. The observation area is put on the active part of the stage, cooled with a nitrogen flux, and a $2\,\mathrm{mm}$ hole in the active part allows white light observation under the microscope in transmission mode. The temperature is calibrated before each set of experiment by replacing the capillary by two successive capillaries filled with pure chemicals whose melting points had previously been calibrated, as described in Ref.~\onlinecite{Ragueneau2022}. The two chemicals that were used in this study are pure water and octane, whose melting point are $273.15$ and $217.55\,\mathrm{K}$, respectively.

\begin{table*}
    \centering
    \caption{Self-diffusion datasets that were used from the literature.}
    \begin{tabular}{cccccc}
        First author and reference        &   Year    &   Pressure range (MPa)    &   Temperature range (K)   &     Accuracy ($\%$)     &   Number of data    \\
        \hline
        Woolf \cite{Woolf1974,Woolf1975}  &1974-1975  &     51.3 - 51.8           &     277.2 - 298.2         &             0.8         &           3         \\
                                          &           &     90.4 - 108.2          &     283.2 - 318.2         &             0.8         &           8         \\
        \hline
        Angell \cite{Angell1976}          &     1976  &  47.48 - 51.45            &     268.16 - 275.36       &             3           &           3         \\
                                          &           &  100.78 - 109.80          &     263.16 - 275.36       &             3-4         &           5         \\
                                          &           &  149.43 - 150.81          &     258.16 - 268.16       &             3-5         &           3         \\
        \hline
        Krynicki \cite{Krynicki1978}      &     1978  &   50                      &     298.2 - 498.2         &             5           &           10        \\
                                          &           &   90 - 110                &     298.2 - 498.2         &             5           &           20        \\
                                          &           &   150                     &     298.2 - 498.2         &             5           &           10        \\
        \hline
        Harris \cite{Harris1980}          &     1980  & 50.1 - 53                 &     277.15 - 333.15       &             2           &           5         \\
                                          &           & 100 - 104                 &     277.15 - 333.15       &             2           &           6         \\
                                          &           & 149.5 - 150.9             &     277.15 - 318.15       &             2           &           5         \\
        \hline
        Easteal \cite{Easteal1984}        &     1984  &         51                &     323.15                &             0.8        &            1        \\
        \hline
        Baker \cite{Baker1985}            &     1985  &         50                &     298.16                &             5           &           1         \\
                                          &           &         100               &     298.16                &             5           &           1         \\
                                          &           &         150               &     298.16                &             5           &           1         \\
        \hline
        Prielmeier \cite{Prielmeier1988}  &     1988  &         50                &     243 - 273             &         3.5 - 4.4       &           8         \\
                                          &           &         100               &     238 - 273             &         3.4 - 4.4       &           9         \\
                                          &           &         150               &     228 - 273             &         3.4 - 4.7       &           11        \\
        \hline
        Harris \cite{Harris1997}          &     1997  &         50 - 52.5         &     268.16 - 298.2        &             0.8         &           3         \\
                                          &           &         100 - 102.5       &     263.17 - 298.21       &             0.8         &           4         \\
                                          &           &         150.5 - 151.0     &     263.19 - 298.16       &             0.8         &           5         \\
        \hline
    \end{tabular}
    \label{tab:DataDiffusion}
\end{table*}

The pressure is changed with a manual screw pump (HIP) able to reach $200\,\mathrm{MPa}$. A separating piston (Top Industrie) was placed between the pump and the capillary to prevent polystyrene particles to invade the high pressure equipment. Two pressure sensors were placed on each side of the piston (HBM P3TCP/3000BAR for pressures above $100\,\mathrm{MPa}$ and Keller PA33X/1000bars for pressures up to $90\,\mathrm{MPa}$). They allow measuring the pressure inside the sample and checking that the piston is not in abutment.

Measurements are taken until crystallization occurs, due to heterogeneous nucleation. This is detected when the CCD image stops fluctuating. Crystallization causes aggregation of the colloids, and the sample must then be replaced.

\subsection{Data treatment}

A key parameter to convert the diffusion coeffcient $D$ of the spheres into liquid viscosity is the sphere radius $a$. An imperfect value may introduce a systematic bias in the data. Moreover, the hydrodynamic interaction with the walls introduces a correction to the SER for $D$ of the spheres (the Oseen correction). Fortunately, this correction depends only on the ratio between the geometric parameters of the experiment, and therefore not on the viscosity itself\cite{Dehaoui2015}. As the thermal expansion coefficient of polystyrene and fused silica are small, the error on viscosity and on the Oseen correction due to the change in the capillary size and sphere radius with temperature is smaller than the statistical error bars of our measurements.

Therefore, at each pressure $P$, a constant effective radius for the spheres was used, determined from the accurate literature data at $T_0=293.15\,\mathrm{K}$, $\eta(T_0,P)$, calculated using the International Association for the Properties of Water and Steam (IAPWS) formulation for the viscosity of water\cite{Huber2009}. The sphere diffusion coefficient $D(T,P)$ measured at temperature $T$ and pressure $P$ was thus converted into viscosity using:
\begin{equation}
\eta(T,P)=\frac{D(T_0,P)}{D(T,P)}\frac{T}{T_0}\eta(T_0,P). 
\end{equation}

For each run, $D(T_0,P)$ was obtained as the average over 10 measurements at $T_0$, with a typical uncertainty of $3\%$ ($68\%$ confidence interval). Then, at other temperatures, the measurement was repeated from 3 to 5 times. The precision on the temperature is 0.13~K, leading to a possible higher uncertainty at low temperatures (where the viscosity varies faster with temperature) than at high temperatures. As described in previous papers\cite{Dehaoui2015, Ragueneau2022}, to take this into account, we fitted each independent run by a Speedy-Angell law $\eta(T)=\eta_0\left(\frac{T}{T_s}-1\right)^{-\gamma}$ and added the contribution of temperature to the global errorbar. The overall uncertainty with a coverage factor $k=1$ lies between 3 and $3.2\%$ (see details in Appendix \ref{App:Data}, Table \ref{tab:DataViscosite}).

\subsection{Set of viscosity data\label{sec:etaset}}

To complement our two data sets for supercooled water under pressure, we use our DDM measurements at atmospheric pressure at 239.15~K \cite{Dehaoui2015} and from 240.15 to 249.15~K \cite{Ragueneau2022} together with the selection of literature data described in Ref.~\onlinecite{Ragueneau2022}. For stable water under pressure, we use the IAPWS formulation for the viscosity of water\cite{Huber2009} to calculate values every 5K from 273.15K to a maximum temperature depending on pressure, and at least equal to 593.16\,K. The uncertainty on those values is the one provided in Fig. 24 of Ref.~\onlinecite{Huber2009}, divided by 2 to cover the $68\%$ confidence interval.

We treated experimental data as belonging to the same isobar as long as their pressures differ by less than $\pm 3\,\mathrm{MPa}$. This allowed considering as a single isobar the data measured under $33\,\mathrm{MPa}$ of Ref.~\onlinecite{Singh2017} and the present data measured under $30\,\mathrm{MPa}$. This introduces a negligible error on viscosity.

\begin{table}
    \centering
    \caption{Tolerance on pressure difference to consider self-diffusion data as isobars.}
    \begin{tabular}{lrrr}
        $P$ (MPa) & $dP$ (MPa)\\
        \hline
        0.1       &      3      \\
        50        &      3      \\
        100       &      10     \\
        150       &      1      \\
        \hline
    \end{tabular}
    \label{tab:IsobarsDiffusion}
\end{table}

\subsection{Set of self-diffusion data\label{sec:Dset}}

To study the Stokes-Einstein ratio, we used literature data for the self-diffusion coefficient of water. In the supercooled liquid under pressure, Prielmeier~\textit{et al.}\cite{Prielmeier1988} cover the broadest range, reaching $228\,\mathrm{K}$ at $150\,\mathrm{MPa}$. It is therefore natural to select self-diffusion values from this source to study the SER in supercooled water under pressure.

In the stable liquid at 0.1\,MPa, we used the same set of data as before\cite{Dehaoui2015}. But in supercooled water several authors\cite{Prielmeier1988,Gillen1972,Price1999} provide data which are mutually inconsistent with each other (see the comparison given in Appendix \ref{App:Diffusion}). Unfortunately, we do not have a compelling argument to select which dataset is the most reliable. 

For consistency with higher pressure, we decided to rely on the data of Prielmeier~\textit{et al.}\cite{Prielmeier1988}. However, at ambient pressure, this data set reaches $252\,\mathrm{K}$ only. To benefit from the lower temperature reached by Gillen~\textit{et al.}\cite{Gillen1972} and Price~\textit{et al.}\cite{Price1999} ($242.5\,\mathrm{K}$ and $237.8\,\mathrm{K}$, respectively), we used the diffusivity data given in these references, but corrected their respective temperature scales with a linear function to match the data of Prielmeier~\textit{et al.}\cite{Prielmeier1988} in the overlapping temperature range (see Appendix~\ref{App:Diffusion} for more details). The sensitivity of the results on this choice will be discussed in Section~\ref{sec:SER}.

Under higher pressures we used the data summarized in table \ref{tab:DataDiffusion}. Su\'arez-Iglesias \cite{Suarez2015} collected all those data in the tables provided in their SI but made some typos regarding Woolf's data. To create our data files, we copied-pasted the tables of Su\'arez-Iglesias and corrected the typos.

Following Su\'arez-Iglesias, we discarded the data of Krynicki\cite{Krynicki1978} below 298.16K. The data of Woolf \cite{Woolf1974, Woolf1975} and Easteal \cite{Easteal1984} were corrected as described by Mills \cite{Mills1973} to deduce the self-diffusion of water from the diffusion of an isotope in water they provide.

As none of those authors specifies the confidence interval corresponding to their uncertainties, we assumed they relate to a $68\%$ confidence interval. As already described in Ref.~\onlinecite{Ragueneau2022}, Prielmeier's uncertainties were corrected to propagate the uncertainty on the temperature, that was not taken into account in their paper. The uncertainties we considered for Angell's data \cite{Angell1976} linearly increase from $3\%$ at -5°C to $6\%$ at -20°C, and remain equal to $3\%$ above -5°C.

To gather data sets obtained at nearly equal pressures, we allowed data taken between $P-\Delta P$ and $P+\Delta P$ to be considered as belonging to the same isobar at $P$; Table~\ref{tab:IsobarsDiffusion} gives the values used for $\Delta P$. In particular, the relatively large $\Delta P$ at 100~MPa allows averaging together the sets of data of Krynicki \cite{Krynicki1978} at 90 and 110~MPa. 

The self-diffusion data were fitted along each isotherm to interpolate the self-diffusion coefficient at the same pressure and temperature as the available viscosity data. The best fitting function we found was $\ln(D) = f[\ln(T)]$, $f$ being a 6-th degree polynomial. 6 fitting functions were computed, one for each of the 6 isobars. All the points presented in Table \ref{tab:DataDiffusion} were used. The uncertainties on the obtained fitted values were assigned equal to the uncertainty of the closest-temperature experimental point on the same isobar.

\section{Results\label{sec:results}}

\begin{figure}
\centering
\includegraphics[width=.8\linewidth]{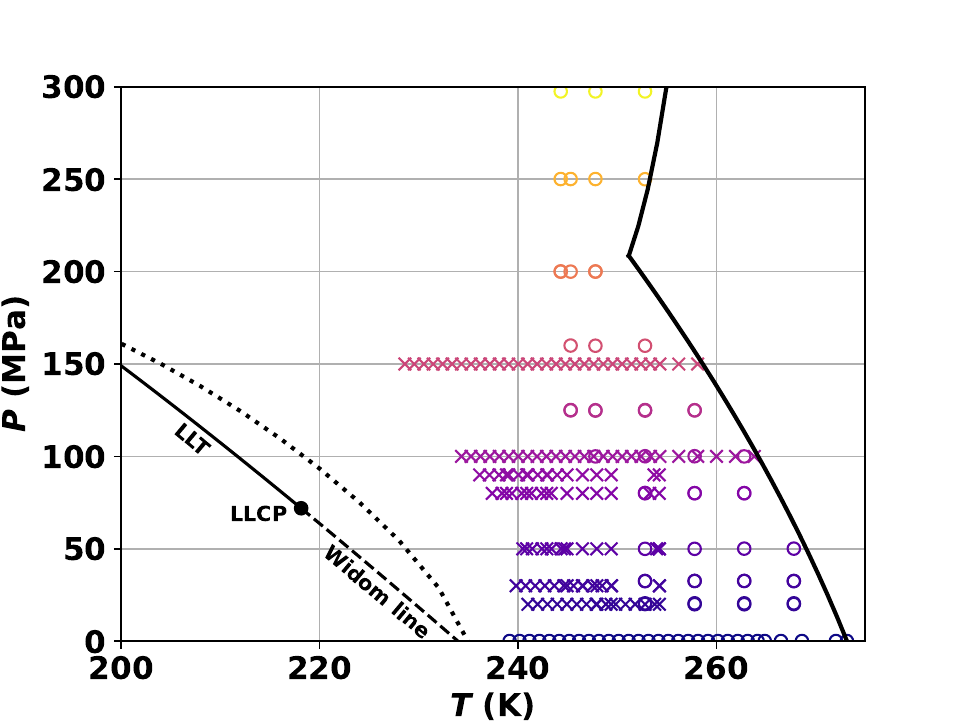}
\caption{Experimental phase diagram of water. The solid-liquid coexistence line is computed following IAPWS release R14-08 \cite{IAPWS2011}. The homogeneous nucleation line (dotted line) is computed following IAPWS guideline G12-15 \cite{IAPWS2015}. The liquid-liquid coexistence line, critical point and Widom line are estimated thanks to a two-state model fitted on a large amount of experimental data \cite{Caupin2019}. Circles show previously existing measurements of the viscosity in supercooled water \cite{Dehaoui2015, Ragueneau2022, Singh2017}. Crosses show present data.}
\label{fig:Domaine}
\end{figure}

The present results obtained with DDM extend our previous data obtained with Poiseuille flow \cite{Singh2017} to lower temperatures at pressures up to $150\,\mathrm{MPa}$, as illustrated in Fig.~\ref{fig:Domaine}.

\begin{figure}
\centering
\includegraphics[width=.8\linewidth]{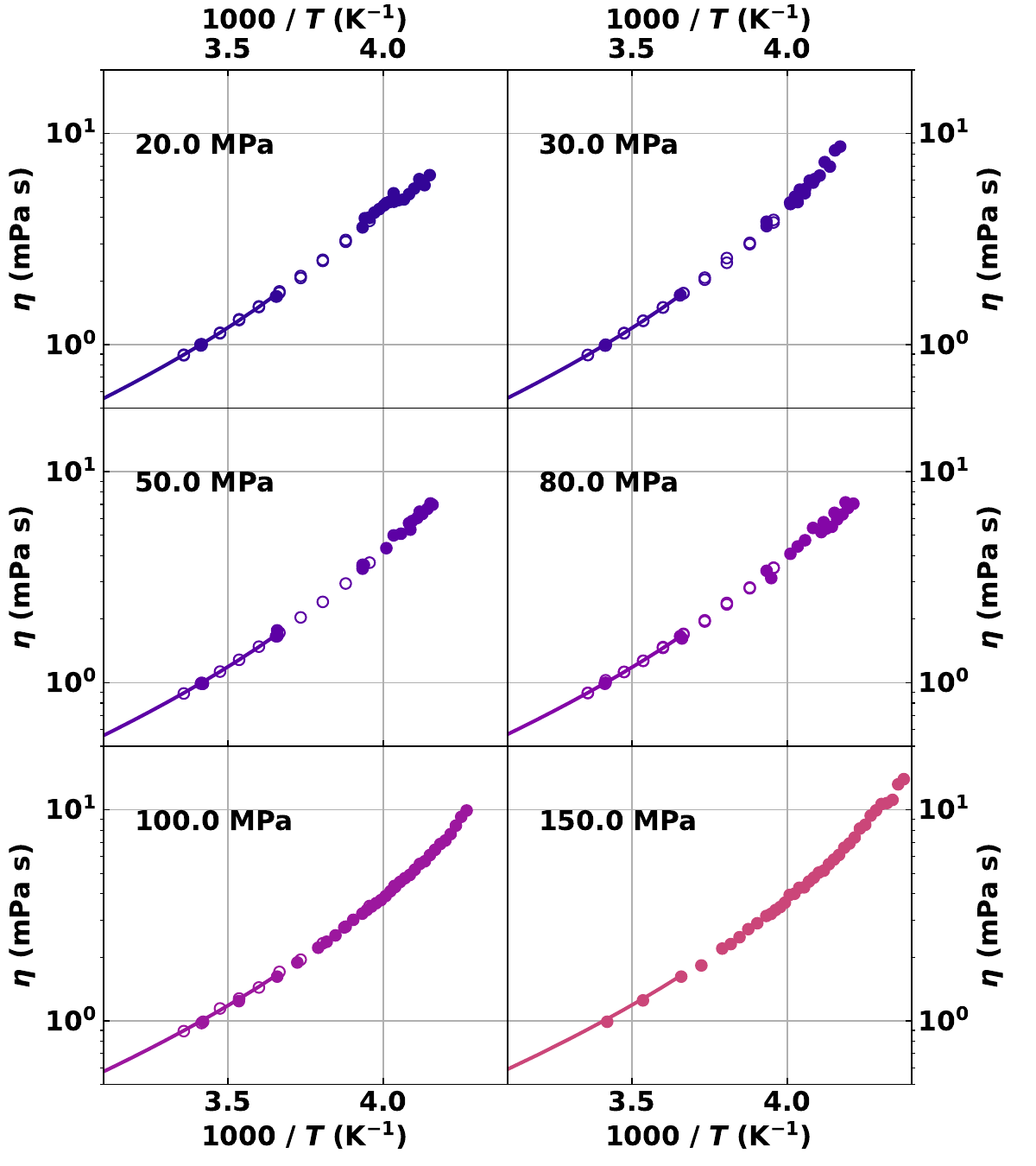}
\caption{Viscosity of water along isobars in Arrhenius representation. Open symbols: Poiseuille measurements \cite{Singh2017}. Full symbols: DDM measurements (this work). Curves: IAPWS formulation \cite{Huber2009}.}
\label{fig:Arrhenius}
\end{figure}

\begin{figure*}
\centering
\includegraphics[width=16cm]{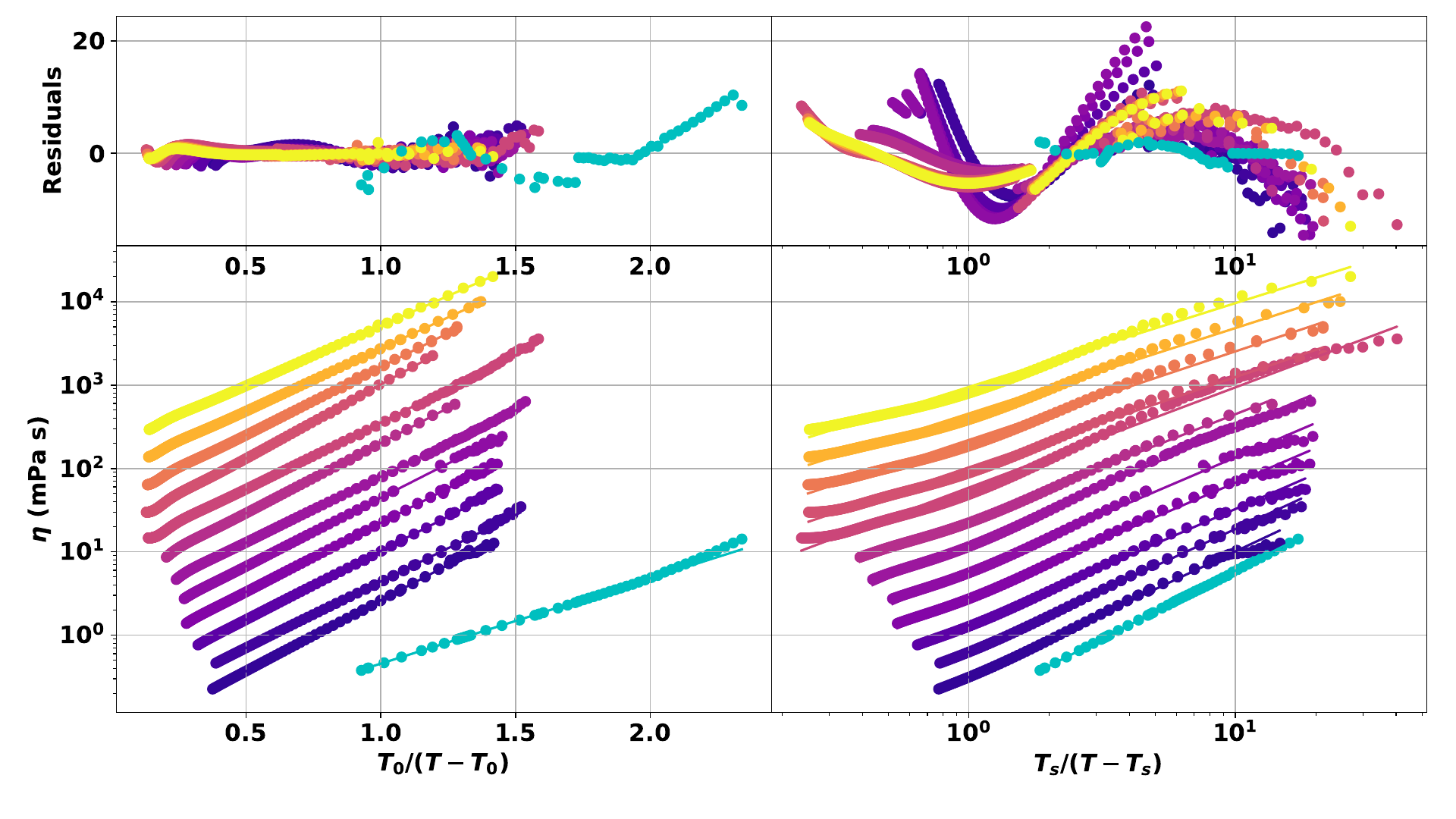}
\caption{Fits of the viscosity data for each isobar by a VTF law (left panel) and a Speedy-Angell law (right panel). Cyan circles: 0.1\,MPa; other circles, from bottom to top: 20, 30, 50, 80, 90, 100, 125, 150, 160, 200, 250, 297.5\,MPa. Data on successive isobars are multiplied by 2 for clarity. The top panels show the reduced residuals (residuals divided by $1-\sigma$ error bar).  While the Speedy-Angell law best describes the atmospheric pressure data, it fails to describe all the higher pressures ones. In contrast, VTF law performs well under pressure: the higher the pressure, the better the fit.}
\label{fig:VTF}
\end{figure*}

\subsection{Temperature variation of viscosity along isobars\label{sec:etaP}}

The viscosity of water along isobars is plotted in Fig.~\ref{fig:Arrhenius} in an Arrhenius representation. The present results are in excellent agreement with previous measurements \cite{Singh2017} and the IAPWS formulation~\cite{Huber2009}, and extend to far lower temperatures. At all pressures, water behaves as a fragile glassformer: its viscosity cannot be described by a simple Arrhenius law $\eta \propto \exp \left( -E_a / k_\mathrm{B} T \right)$, with constant activation energy $E_a$. This confirms the findings of our previous work~\cite{Singh2017} covering a larger pressure range but reaching a smaller degree of supercooling.

It was already noted in Refs.~\onlinecite{Dehaoui2015} and \onlinecite{Ragueneau2022} that data at ambient pressure is best represented by the Speedy-Angell law:
\begin{equation}
    \eta(P=0.1\,\mathrm{MPa},T) = \eta_0\left(\frac{T}{T_s}-1\right)^{-\gamma} \, .
    \label{eq:SpeedyAngell}
\end{equation}
At all elevated pressures however, the Vogel-Tamman-Fulcher law (VTF, eq.~\ref{eq:VTF}):
\begin{equation}
    \eta(P,T) = \eta_0(P)\exp\left(-\frac{B(P)}{T-T_0(P)}\right) \, ,
    \label{eq:VTF}
\end{equation}
a popular representation of the temperature dependence of viscosity in fragile glassformers, performs better (see Fig.~\ref{fig:VTF}). It is noteworthy that a modest increase in pressure, by $20\,\mathrm{MPa}$ only, is sufficient to induce a significant qualitative change in the temperature dependence of viscosity. The values of the fitting coefficients of all isotherms by both equations are provided in Appendix~\ref{app:FitCoef}.

\begin{figure}
\centering
\includegraphics[width=.8\linewidth]{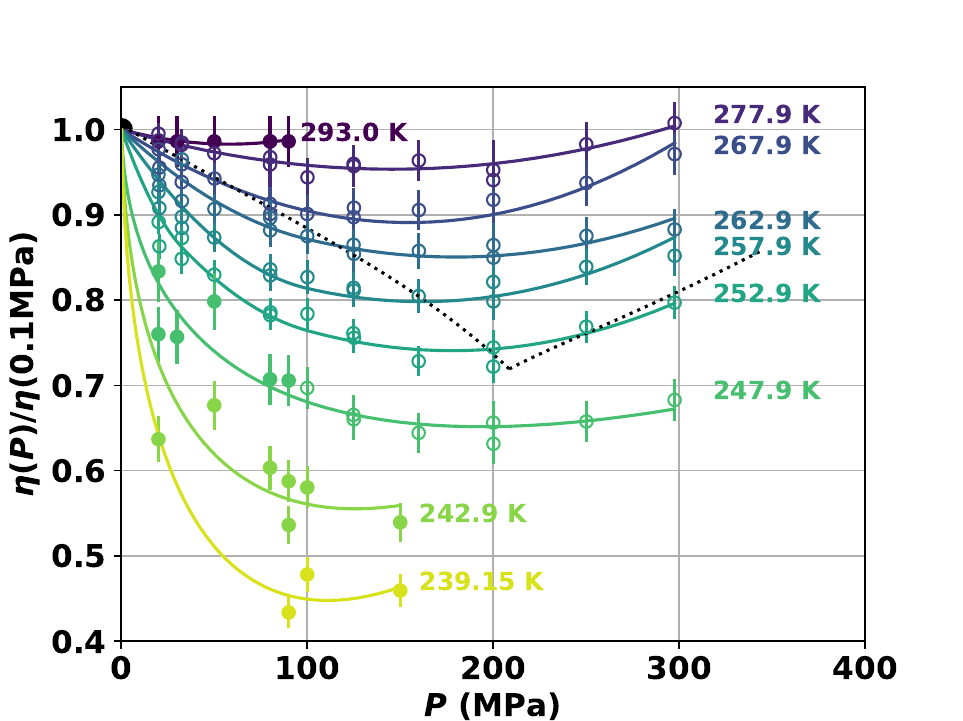}
\caption{Viscosity of water along isotherms. The values are normalized by the value at atmospheric pressure\cite{Dehaoui2015,Ragueneau2022}, which makes all isotherms go through the point (0.1\,MPa, 1). Open symbols: Poiseuille measurements \cite{Singh2017}. Full symbols: DDM measurements (this work). For clarity, not all isotherms are shown. Dashed line: Melting line. Plain colored lines are guides to the eye. The error bars indicate a $68\%$ confidence interval.}
\label{fig:eta_Isothermes}
\end{figure}

\subsection{Pressure variation of viscosity along isotherms\label{sec:etaT}}

Figure~\ref{fig:eta_Isothermes} displays the viscosity of water as a function of pressure along different isotherms. To emphasize the pressure dependence only, along each isotherm, the ratio between the viscosity under pressure and its value at ambient pressure is shown, which makes all curves go through the point (0.1\,MPa, 1). Note that, in this representation, the isotherms cannot be plotted at temperatures lower than 239.15\,K since the viscosity at atmospheric pressure was never measured below that temperature.

The anomalous decrease of viscosity with pressure, discovered in 1884~\cite{Roentgen1884,Warburg1884}, and confirmed in the supercooled region by our previous work~\cite{Singh2017}, is seen here to get even more pronounced at larger supercooling. Our new data show that at 239.15\,K the viscosity is divided by more than 2 between 0 and $150\,\mathrm{MPa}$; it is likely that a minimum will be eventually reached, but it lies beyond the highest pressure accessible with our experiment.

\subsection{Stokes-Einstein relation\label{sec:SER}}

\begin{figure}
\centering
\includegraphics[width=.8\linewidth]{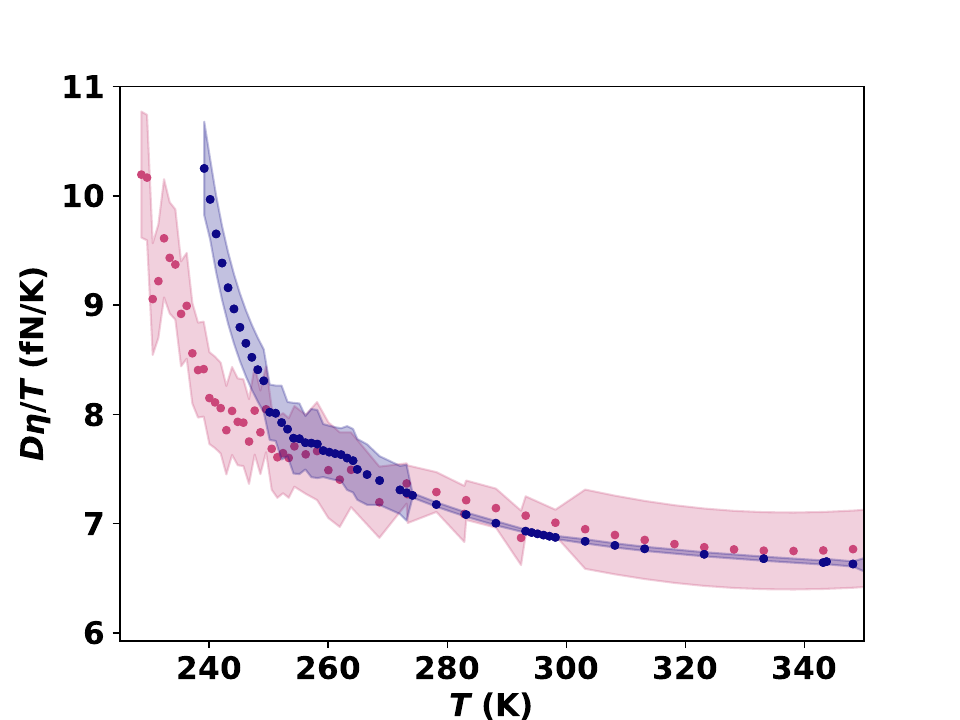}
\caption{Stokes-Einstein ratio in water along two isobars at $0.1\,\mathrm{MPa}$ (purple symbols) and $150\,\mathrm{MPa}$ (pink symbols). The shaded area shows the error band.}
\label{fig:SE_BarresDErreur}
\end{figure}

\begin{figure}
\centering
\includegraphics[width=.8\linewidth]{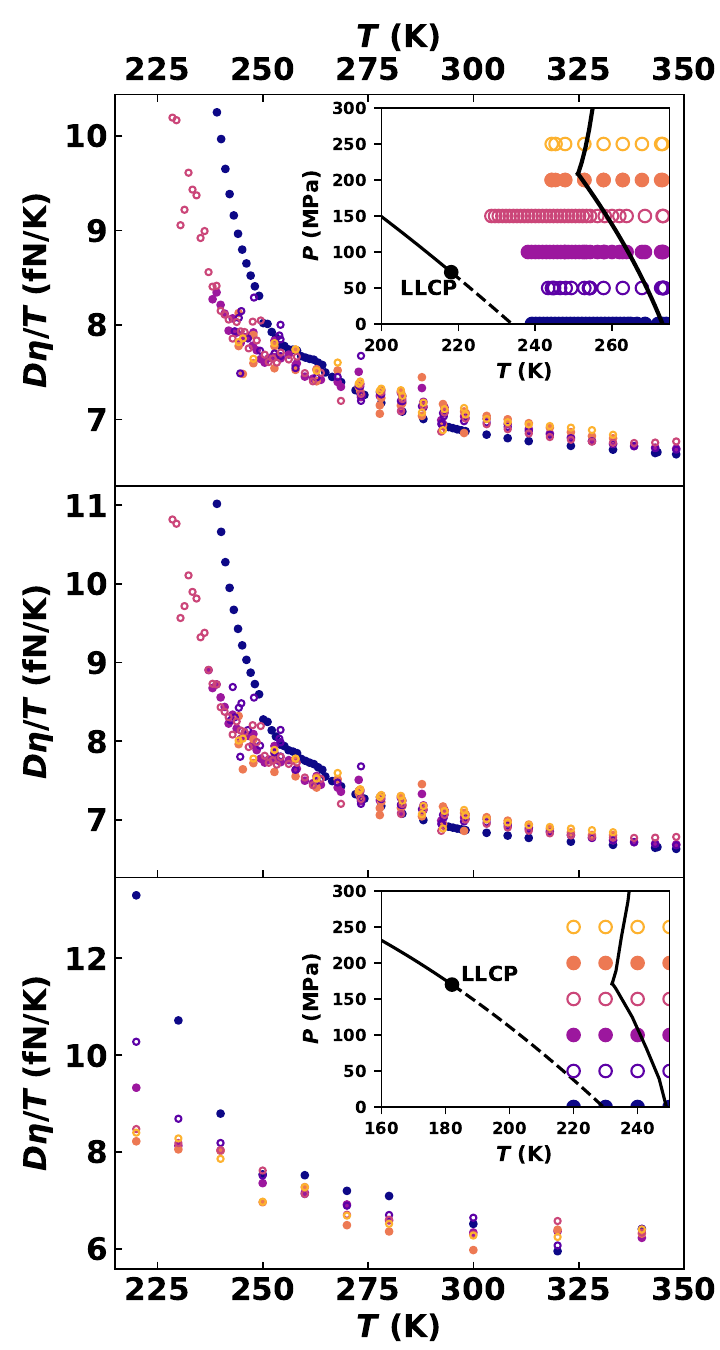}
\caption{Stokes-Einstein ratio in water along 6 isobars (see phase diagrams in the insets) as functions of temperature. The LLT, LLCP, and Widom line shown in the experimental phase diagram are computed from a two-state model~\cite{Caupin2019} as in Fig.~\ref{fig:Domaine}. Top panel: experimental data. The temperatures of all data for self-diffusion at $0.1\,\mathrm{MPa}$ in the supercooled area were rescaled as explained in Appendix~\ref{App:Diffusion} to match the $0.1\,\mathrm{MPa}$ data of Ref.~\onlinecite{Prielmeier1988}. Error bars (not shown for clarity) are comparable to Fig.~\ref{fig:SE_BarresDErreur}. Middle panel: same as top panel, except that the temperatures of all data were rescaled to match the $0.1\,\mathrm{MPa}$ data of Ref.~\onlinecite{Price1999}. Bottom panel: molecular dynamics simulations of Dubey\cite{Dubey2019} with the TIP4P/2005 water model.
}
\label{fig:SE_Isobares}
\end{figure}

Having obtained viscosity data in a previously uncharted region of the temperature-pressure plane, we can now study the effect of pressure on the SER, combining data on viscosity (Sections~\ref{sec:etaset} and \ref{sec:etaP}) and self-diffusion (Section~\ref{sec:Dset}). Figure~\ref{fig:SE_BarresDErreur} shows the Stokes-Einstein ratio at 0.1 and $150\,\mathrm{MPa}$, with the corresponding error bands, which are more apparent than for the separate quantities $D$ and $\eta$, due to the more modest variation of the ratio with temperature. Nevertheless, the uncertainty is sufficiently small to capture the difference between the two isobars.

We now plot in the top panel of Fig.~\ref{fig:SE_Isobares} the Stokes-Einstein ratio along all measured isobars. The first striking feature, already noticed before, is that at high temperature all curves reach a similar constant value. If we convert it into a hydrodynamic radius $R_\mathrm{h}=k_\mathrm{B}T/(4\pi\eta D_\mathrm{s})$ (the factor 4 corresponds to a full slip boundary condition), we get $2R_\mathrm{h}\simeq 0.32\,\mathrm{nm}$, close to the size of a water molecule. This is rather puzzling, as hydrodynamic laws are not expected to hold at the molecular scale. The second striking feature is the increase of the Stokes-Einstein ratio upon cooling. This violation of SER, observed at $0.1\,\mathrm{MPa}$\cite{Dehaoui2015}, is thus confirmed at all pressure up to $150\,\mathrm{MPa}$, where it reaches up to 50\%. In this pressure range, the glass transition temperature $T_g$ remains close to $140\,\mathrm{K}$\cite{AmannWinkel2013}: the SER always starts being violated at around $2T_g$.

Interestingly, within experimental uncertainty, the SER violation starts at higher temperature at $0.1\,\mathrm{MPa}$, while the other isobars cannot be distinguished from each other in the temperature range we could cover (see also Supplementary Fig.~S1 which plots $D\eta/T$ along several low temperature isotherms to further illustrate this point). To see if this is an effect of the temperature correction applied to $D_\mathrm{s}$ data at ambient pressure (Section~\ref{sec:Dset}), we plot in the middle panel of Fig.~\ref{fig:SE_Isobares} the Stokes-Einstein ratio obtained with an alternative correction, now rescaling the temperatures of Prielmeier~\textit{et al.} and Gillen~\textit{et al.} to those of Price~\textit{et al.}. All curves are shifted in temperature, so that the difference in SER violation between ambient pressure and all other pressures remains (see also Supplementary Fig.~S1). We also note in that case that the ratio reaches a higher value at low temperature.

\section{Discussion\label{sec:disc}}

To investigate possible implications of our findings in the debate about the putative liquid-liquid transition in supercooled water, we compare with molecular dynamics simulations performed with the TIP4P/2005 model for water. This force field is one of the most accurate~\cite{Vega2011}. It exhibits a LLCP, whose location is estimated to $T_\mathrm{c}=182\,\mathrm{K}$ and $P_\mathrm{c}=170\,\mathrm{MPa}$ with a two-state model fitted on simulation data\cite{Biddle2017}, or at $T_\mathrm{c}=172\,\mathrm{K}$ and $P_\mathrm{c}=186\,\mathrm{MPa}$ with histogram reweighting\cite{Debenedetti2020}. Self-diffusion and viscosity values have been obtained in the stable and supercooled region in a broad pressure range~\cite{Montero2018,Dubey2019}, and show a good agreement with experimental data.

\begin{table}
    \centering
    \caption{Various predictions for the LLCP temperature $T_\mathrm{c}$ and pressure $P_\mathrm{c}$ in real water, based on experimental data only.}
    \begin{tabular}{ccc}
Source            				& $T_\mathrm{c}$ (K)	&  $P_\mathrm{c}$   (MPa)     \\
\hline
Ref.~\onlinecite{Holten2012}\footnote{mean-field equation of state; the Authors mention that ``the optimum locations of the critical point form a narrow band in the $P$--$T$ diagram, which extends approximately from $222\,\mathrm{K}$ and $50\,\mathrm{MPa}$ to $240\,\mathrm{K}$ and $30\,\mathrm{MPa}$}	
								&228.2					&0\\
Ref.~\onlinecite{Holten2012}\footnote{crossover equation of state}
								&227.4					&13.5\\
Ref.~\onlinecite{Duska2020}		&220.9					&54.2\\
Ref.~\onlinecite{Caupin2019}	&218.1					&71.9\\
Ref.~\onlinecite{Mishima1998}	&220					&100\\
Ref.~\onlinecite{Mishima2023}	&207					&105\\
Ref.~\onlinecite{Shi2020}		&184					&173\\
Ref.~\onlinecite{Bachler2021}	&180					&200\\
\hline
    \end{tabular}
    \label{tab:Pc}
\end{table}

The Stokes-Einstein ratio simulated by Dubey~\textit{et al.}\cite{Dubey2019} is displayed in Fig.~\ref{fig:SE_Isobares}, bottom panel. As pointed out in Ref.~\onlinecite{Dubey2019}, below $250\,\mathrm{K}$, $D\eta/T$ decreases as the pressure increases for a given temperature. The simulations suggest that, within the considered temperature range, $D\eta/T$ becomes nearly independent of pressure, above a pressure close to the LLCP pressure, see Fig.~\ref{fig:SE_Isobares}. A similar trend is observed in experiments (Fig.~\ref{fig:SE_Isobares}): at all temperatures below $245\,\mathrm{K}$, $D\eta/T$  is highest at $0.1\,\mathrm{MPa}$; $D\eta/T$ also seems to become pressure independent above $50\,\mathrm{MPa}$, but supporting data only consists of a few points over a limited temperature range (from 240 to $245\,\mathrm{K}$, see Figs.~\ref{fig:SE_Isobares} and S1). Several estimates have been proposed for a LLCP in real water, with pressures ranging from 0 to $200\,\mathrm{MPa}$ (see Table~\ref{tab:Pc}). Our data are compatible with a LLCP at positive pressure in real water, but further studies, measuring the pressure dependence of $D\eta/T$ at even lower temperatures, closer to the predicted Widom line as in the simulations (see Fig.~\ref{fig:SE_Isobares}, insets), are needed to decide between the various LLCP predictions.

\appendix

\section{Raw measurements of the viscosity of pure light water under high pressure}
\label{App:Data}

\begin{longtable}{|c|c|c|c|}
\caption{Raw values and $1-\sigma$ error bar for the viscosity of $\rm{H_2O}$ under pressure.}
\\
\hline
P (MPa)   &   T (K)   &   $\eta$ ($\mathrm{mPa\,s}$)   &   $d\eta$ ($\mathrm{mPa\,s}$)   \\ 
\hline
\endhead
\hline
\endfoot
\hline
\multirow{7}{*}{20.0}   &   241.00   &   6.4   &   0.3   \\ 
   &   241.98   &   5.69   &   0.19   \\ 
   &   242.97   &   6.1   &   0.2   \\ 
   &   243.95   &   5.48   &   0.18   \\ 
   &   244.93   &   5.16   &   0.17   \\ 
   &   245.92   &   4.87   &   0.16   \\ 
   &   246.90   &   4.84   &   0.16   \\ 
\multirow{14}{*}{20.0}      &   247.88   &   4.75   &   0.15   \\ 
   &   247.94   &   5.21   &   0.17   \\ 
   &   248.86   &   4.73   &   0.15   \\ 
   &   249.39   &   4.68   &   0.15   \\ 
   &   249.85   &   4.57   &   0.15   \\ 
   &   250.83   &   4.38   &   0.14   \\ 
   &   251.81   &   4.22   &   0.13   \\ 
   &   252.80   &   4.00   &   0.13   \\ 
   &   253.78   &   3.97   &   0.13   \\ 
   &   254.24   &   3.59   &   0.12   \\ 
   &   273.44   &   1.69   &   0.06   \\ 
   &   273.62   &   1.69   &   0.06   \\ 
   &   293.00   &   0.99   &   0.03   \\ 
   &   293.11   &   1.00   &   0.04   \\ 
\hline
\multirow{13}{*}{30.0}   &   239.81   &   8.7   &   0.3   \\ 
   &   240.77   &   8.3   &   0.3   \\ 
   &   241.73   &   7.0   &   0.3   \\ 
   &   242.70   &   7.3   &   0.3   \\ 
   &   243.66   &   6.3   &   0.2   \\ 
   &   244.62   &   6.08   &   0.19   \\ 
   &   244.94   &   5.85   &   0.19   \\ 
   &   245.59   &   5.98   &   0.19   \\ 
   &   246.49   &   5.45   &   0.17   \\ 
   &   246.55   &   5.21   &   0.17   \\ 
   &   247.52   &   5.42   &   0.17   \\ 
   &   247.94   &   4.73   &   0.15   \\ 
   &   248.48   &   5.02   &   0.16   \\ 
\multirow{8}{*}{30.0}   &   249.39   &   4.63   &   0.15   \\ 
   &   249.44   &   4.71   &   0.15   \\ 
   &   254.24   &   3.64   &   0.12   \\ 
   &   254.26   &   3.82   &   0.12   \\ 
   &   273.54   &   1.72   &   0.06   \\ 
   &   273.62   &   1.71   &   0.06   \\ 
   &   292.81   &   0.99   &   0.03   \\ 
   &   293.00   &   0.99   &   0.03   \\ 
\hline
\multirow{19}{*}{50.0}   &   240.48   &   7.0   &   0.3   \\ 
   &   240.87   &   7.1   &   0.3   \\ 
   &   241.44   &   6.7   &   0.3   \\ 
   &   242.41   &   6.3   &   0.2   \\ 
   &   242.90   &   6.5   &   0.2   \\ 
   &   243.37   &   6.03   &   0.19   \\ 
   &   244.34   &   5.83   &   0.18   \\ 
   &   244.68   &   5.31   &   0.17   \\ 
   &   244.94   &   5.70   &   0.18   \\ 
   &   246.49   &   5.08   &   0.16   \\ 
   &   247.94   &   4.99   &   0.16   \\ 
   &   249.39   &   4.34   &   0.14   \\ 
   &   253.97   &   3.61   &   0.12   \\ 
   &   254.24   &   3.55   &   0.11   \\ 
   &   254.24   &   3.47   &   0.11   \\ 
   &   254.24   &   3.62   &   0.12   \\ 
   &   273.25   &   1.68   &   0.06   \\ 
   &   273.34   &   1.66   &   0.06   \\ 
   &   273.34   &   1.77   &   0.06   \\ 
\multirow{5}{*}{50.0}   &   273.62   &   1.66   &   0.06   \\ 
   &   292.45   &   0.99   &   0.03   \\ 
   &   292.45   &   0.99   &   0.03   \\ 
   &   292.52   &   0.99   &   0.03   \\ 
   &   293.00   &   0.99   &   0.03   \\ 
\hline
\multirow{20}{*}{80.1}   &   237.39   &   7.0   &   0.3   \\ 
   &   238.38   &   6.7   &   0.3   \\ 
   &   238.83   &   7.2   &   0.3   \\ 
   &   239.38   &   6.3   &   0.2   \\ 
   &   240.37   &   5.94   &   0.19   \\ 
   &   240.87   &   6.4   &   0.2   \\ 
   &   241.36   &   5.48   &   0.17   \\ 
   &   242.36   &   5.38   &   0.17   \\ 
   &   242.90   &   5.76   &   0.18   \\ 
   &   243.35   &   5.17   &   0.16   \\ 
   &   244.94   &   5.41   &   0.17   \\ 
   &   246.49   &   4.73   &   0.15   \\ 
   &   247.94   &   4.42   &   0.14   \\ 
   &   249.39   &   4.08   &   0.13   \\ 
   &   253.28   &   3.13   &   0.10   \\ 
   &   254.24   &   3.39   &   0.11   \\ 
   &   273.15   &   1.62   &   0.05   \\ 
   &   273.62   &   1.66   &   0.06   \\ 
   &   293.00   &   0.99   &   0.03   \\ 
   &   293.02   &   0.99   &   0.03   \\ 
\hline
\multirow{2}{*}{90.0}   &   236.13   &   7.6   &   0.3   \\ 
   &   237.11   &   6.5   &   0.3   \\ 
\multirow{20}{*}{90.0}   &   238.08   &   6.8   &   0.3   \\ 
   &   238.83   &   7.0   &   0.3   \\ 
   &   239.06   &   6.2   &   0.2   \\ 
   &   240.04   &   6.2   &   0.2   \\ 
   &   240.87   &   6.2   &   0.2   \\ 
   &   241.01   &   5.68   &   0.18   \\ 
   &   241.99   &   5.46   &   0.17   \\ 
   &   242.90   &   5.61   &   0.18   \\ 
   &   242.97   &   5.12   &   0.16   \\ 
   &   243.94   &   5.03   &   0.16   \\ 
   &   244.94   &   5.05   &   0.16   \\ 
   &   246.49   &   4.70   &   0.15   \\ 
   &   247.94   &   4.41   &   0.14   \\ 
   &   249.39   &   4.18   &   0.13   \\ 
   &   253.71   &   3.20   &   0.10   \\ 
   &   254.24   &   3.41   &   0.11   \\ 
   &   273.25   &   1.67   &   0.06   \\ 
   &   273.62   &   1.65   &   0.05   \\ 
   &   292.78   &   0.99   &   0.03   \\ 
   &   293.00   &   0.99   &   0.03   \\ 
\hline
\multirow{7}{*}{100.0}   &   234.30   &   9.9   &   0.4   \\ 
   &   235.26   &   9.3   &   0.3   \\ 
   &   236.21   &   8.4   &   0.3   \\ 
   &   237.16   &   7.7   &   0.3   \\ 
   &   238.11   &   7.2   &   0.3   \\ 
   &   239.06   &   6.9   &   0.3   \\ 
   &   240.02   &   6.5   &   0.3   \\ 
\multirow{24}{*}{100.0}   &   240.97   &   6.10   &   0.19   \\ 
   &   241.92   &   5.70   &   0.18   \\ 
   &   242.87   &   5.54   &   0.18   \\ 
   &   243.83   &   5.20   &   0.17   \\ 
   &   244.78   &   4.91   &   0.16   \\ 
   &   245.73   &   4.74   &   0.15   \\ 
   &   246.68   &   4.55   &   0.14   \\ 
   &   247.63   &   4.31   &   0.14   \\ 
   &   248.59   &   4.11   &   0.13   \\ 
   &   249.54   &   3.90   &   0.12   \\ 
   &   250.49   &   3.73   &   0.12   \\ 
   &   251.44   &   3.61   &   0.12   \\ 
   &   252.39   &   3.48   &   0.11   \\ 
   &   253.35   &   3.35   &   0.11   \\ 
   &   254.30   &   3.22   &   0.10   \\ 
   &   256.20   &   3.01   &   0.10   \\ 
   &   258.11   &   2.77   &   0.09   \\ 
   &   260.01   &   2.54   &   0.08   \\ 
   &   261.92   &   2.37   &   0.08   \\ 
   &   263.82   &   2.22   &   0.07   \\ 
   &   268.58   &   1.89   &   0.06   \\ 
   &   273.34   &   1.62   &   0.05   \\ 
   &   282.86   &   1.24   &   0.04   \\ 
   &   292.38   &   0.99   &   0.03   \\ 
\hline
\multirow{3}{*}{150.0}   &   228.59   &   14.0   &   0.5   \\ 
   &   229.54   &   13.2   &   0.5   \\ 
   &   230.50   &   11.1   &   0.4   \\ 
\multirow{27}{*}{150.0}   &   231.45   &   10.8   &   0.4   \\ 
   &   232.40   &   10.6   &   0.4   \\ 
   &   233.35   &   9.9   &   0.4   \\ 
   &   234.30   &   9.4   &   0.3   \\ 
   &   235.26   &   8.5   &   0.3   \\ 
   &   236.21   &   8.2   &   0.3   \\ 
   &   237.16   &   7.4   &   0.3   \\ 
   &   238.11   &   6.9   &   0.3   \\ 
   &   239.06   &   6.6   &   0.3   \\ 
   &   240.02   &   6.11   &   0.19   \\ 
   &   240.97   &   5.81   &   0.18   \\ 
   &   241.92   &   5.52   &   0.18   \\ 
   &   242.87   &   5.15   &   0.16   \\ 
   &   243.83   &   5.04   &   0.16   \\ 
   &   244.78   &   4.77   &   0.15   \\ 
   &   245.73   &   4.57   &   0.15   \\ 
   &   246.68   &   4.29   &   0.14   \\ 
   &   247.63   &   4.27   &   0.14   \\ 
   &   248.59   &   4.00   &   0.13   \\ 
   &   249.54   &   3.95   &   0.13   \\ 
   &   250.49   &   3.63   &   0.12   \\ 
   &   251.44   &   3.46   &   0.11   \\ 
   &   252.39   &   3.35   &   0.11   \\ 
   &   253.35   &   3.21   &   0.10   \\ 
   &   254.30   &   3.14   &   0.10   \\ 
   &   256.20   &   2.90   &   0.09   \\ 
   &   258.11   &   2.72   &   0.09   \\ 
\multirow{7}{*}{150.0}   &   260.01   &   2.49   &   0.08   \\ 
   &   261.92   &   2.31   &   0.08   \\ 
   &   263.82   &   2.20   &   0.07   \\ 
   &   268.58   &   1.83   &   0.06   \\ 
   &   273.34   &   1.62   &   0.05   \\ 
   &   282.86   &   1.25   &   0.04   \\ 
   &   292.38   &   0.99   &   0.03   \\ 
\hline
\label{tab:DataViscosite}
\end{longtable}

\section{Recalibration of diffusion coefficients data}
\label{App:Diffusion}
\begin{figure*}
\centering
\includegraphics[width=16cm]{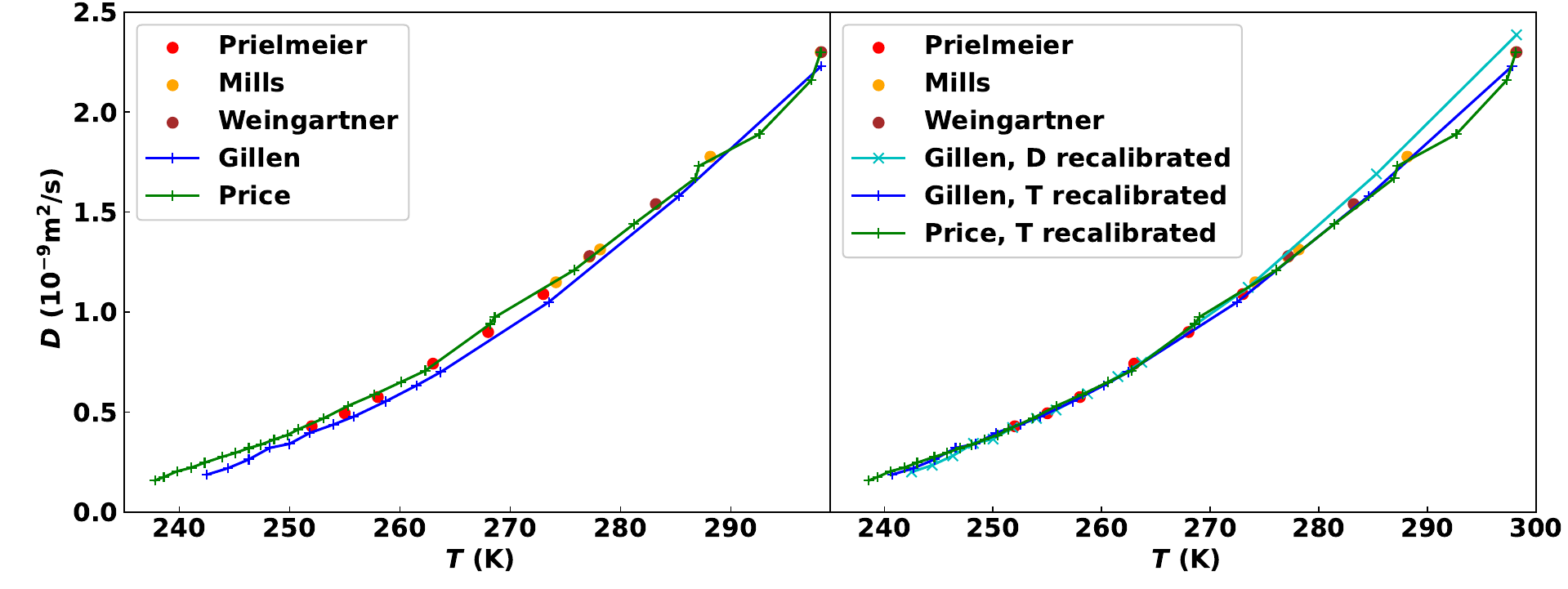}
\caption{Left panel: Self-diffusion coefficients that can be found in the literature: Gillen's data\cite{Gillen1972} systematically underestimate the self-diffusion values compared with Mills, Weingärtner and Prielmeier's data\cite{Mills1973,Weingartner1982,Prielmeier1988}. On the contrary, Price provides the highest values in the supercooled area. Right panel: While multiplying all $D_\mathrm{s}$ values of Gillen by 1.07 as suggested by Prielmeier overestimates the values at positive temperature, we rather recalibrated the temperatures of Price and Gillen by an affine function to make them collapse with the 3 other sets of data.}
\label{fig:CourbesDiffusion}
\end{figure*}

As can be seen in Fig.~\ref{fig:CourbesDiffusion} (left panel), the self-diffusion coefficient measurements available in the literature do not perfectly agree together, in particular in the supercooled area. While Gillen's\cite{Gillen1972} value are systematically lower than other values in the whole temperature range, Price's\cite{Price1999} values systematically lie above Prielmeier's\cite{Prielmeier1988} in the supercooled range. Prielmeier suggests to multiply all Gillen's values by 1.07\cite{Prielmeier1988}. But this tends to overestimate the self diffusion values above 280\,K (right panel, cyan curve). 

Rather than multiplying self-diffusion data by a factor, we find more appropriate to apply a linear rescaling to temperature, because it is able to collapse Gillen's data on other data (right panel, blue curve). Applying also a linear temperature rescaling to Price's data (right panel, green curve) collapses it on Prielmeier's data. Table~\ref{table:PriceGillenCorriges} gives the coefficients that were applied.

\begin{table*}
    \centering
    \caption{Coefficients of the temperature recalibration of Price and Gillen's data: The temperatures $T$ they provided were rescaled as $T_{\rm{rescaled}} = aT+b$.}
    \begin{tabular}{ccc}
                    &       Price\cite{Price1999}     &  Gillen\cite{Gillen1972}       \\
        \hline
            $a$     &           0.9871794871794873    &     1.016931759876859           \\
            $b$ (K) &           3.823076923076883     &     -6.000667008722189          \\
            \hline
    \end{tabular}
    \label{table:PriceGillenCorriges}
\end{table*}

\section{Best-fitting coefficients of the viscosity data by Vogel-Tamman-Fulcher and Speedy-Angell laws}
\label{app:FitCoef}

The coefficients of Eqs.~\ref{eq:SpeedyAngell} and \ref{eq:VTF} that were used to plot Fig.~\ref{fig:VTF} are shown in Tables~\ref{tab:FitVTF} and \ref{tab:FitSA}.

\begin{table*}
    \centering
    \caption{Fitting coefficients of $\eta(P,T)$ by Eq.~\ref{eq:VTF}}
    \begin{tabular}{ccccccc}
        $P$ (MPa)   &  Temperature range (K)  &  Number of points $N$   &   $\eta_0$ ($\mathrm{\mu Pa\,s}$)  &   $-B(P)$ (K) & $T_0(P)$ (K)         &  reduced $\chi^2 = \frac{\chi^2}{N-3}$       \\
        \hline
            0.1     &  239.15 - 348.15        & 49                      &   $43.3 \pm 0.4$    & $394 \pm 3$  &   $167.6 \pm 0.4$    &   18                                         \\
            20      &  241.00 - 518.16        & 91                      &   $26.27 \pm 0.17$  & $553 \pm 3$  &   $141.4 \pm 0.5$    &   1.7                                         \\
            30      &  239.81 - 518.16        & 91                      &   $28.20 \pm 0.18$  & $531 \pm 3$  &   $144.6 \pm 0.5$    &   2.5                                         \\
            50      &  240.48 - 583.16        & 97                      &   $28.77 \pm 0.15$  & $538 \pm 3$  &   $141.7 \pm 0.5$    &   1.3                                         \\
            80      &  237.39 - 643.16        & 115                     &   $30.87 \pm 0.15$  & $533 \pm 3$  &   $139.8 \pm 0.4$    &   0.68                                         \\
            90      &  236.13 - 658.16        & 100                     &   $31.69 \pm 0.15$  & $529 \pm 3$  &   $139.8 \pm 0.5$    &   0.74                                         \\
            100     &  234.30 - 733.16        & 135                     &   $32.4 \pm 0.4$    & $518 \pm 5$  &   $141.9 \pm 0.7$    &   0.56                                         \\
            125     &  245.30 - 813.16        & 133                     &   $32.8 \pm 0.4$    & $533 \pm 5$  &   $137.5 \pm 0.9$    &   0.40                                         \\
            150     &  228.59 - 1168.16       & 217                     &   $33.93 \pm 0.19$  & $520 \pm 3$  &   $140.2 \pm 0.5$    &   1.4                                         \\
            160     &  245.30 - 1168.16       & 192                     &   $33.7 \pm 0.3$    & $544 \pm 4$  &   $133.4 \pm 0.9$    &   0.62                                         \\
            200     &  244.30 - 1168.16       & 203                     &   $38.0 \pm 0.3$    & $514 \pm 4$  &   $137.3 \pm 0.8$    &   0.47                                         \\
            250     &  244.30 - 1168.16       & 193                     &   $42.8 \pm 0.3$    & $487 \pm 4$  &   $141.3 \pm 0.8$    &   0.38                                         \\
            297.5   &  244.30 - 1168.16       & 192                     &   $46.6 \pm 0.3$    & $472 \pm 4$  &   $143.2 \pm 0.9$    &   0.34                                         \\
            \hline
    \end{tabular}
    \label{tab:FitVTF}
\end{table*}

\begin{table*}
    \centering
    \caption{Fitting coefficients of $\eta(P,T)$ by Eq.~\ref{eq:SpeedyAngell}}
    \begin{tabular}{ccccccc}
        $P$ (MPa)   &  Temperature range (K)  &  Number of points $N$   &   $\eta_0$ ($\mathrm{\mu Pa\,s}$)  &   $T_s(P)$ (K)          & $\gamma(P)$            &  reduced $\chi^2 = \frac{\chi^2}{N-3}$       \\
        \hline
            0.1     &  239.15 - 348.15        & 49                      &   $137.4 \pm 0.3$   &     $225.99 \pm 0.14$   &   $1.636 \pm 0.004$   &   1.3                                         \\
            20      &  241.00 - 518.16        & 91                      &   $156.9 \pm 0.4$   &     $225.63 \pm 0.17$   &   $1.509 \pm 0.003$   &   33                                         \\
            30      &  239.81 - 518.16        & 91                      &   $156.7 \pm 0.3$   &     $226.96 \pm 0.15$   &   $1.476 \pm 0.003$   &   26                                         \\
            50      &  240.48 - 583.16        & 97                      &   $163.6 \pm 0.3$   &     $228.02 \pm 0.13$   &   $1.397 \pm 0.003$   &   49                                         \\
            80      &  237.39 - 643.16        & 115                     &   $177.1 \pm 0.3$   &     $225.51 \pm 0.13$   &   $1.377 \pm 0.002$   &   63                                         \\
            90      &  236.13 - 658.16        & 100                     &   $181.7 \pm 0.3$   &     $224.61 \pm 0.13$   &   $1.369 \pm 0.002$   &   77                                         \\
            100     &  234.30 - 733.16        & 135                     &   $196.5 \pm 0.8$   &     $222.7 \pm 0.2$      &   $1.384 \pm 0.006$   &   8.6                                         \\
            125     &  245.30 - 813.16        & 133                     &   $193.3 \pm 0.8$   &     $228.6 \pm 0.3$   &   $1.257 \pm 0.005$   &   8.5                                         \\
            150     &  228.59 - 1168.16       & 217                     &   $229.6 \pm 0.6$   &     $223.06 \pm 0.10$    &   $1.203 \pm 0.003$   &   23                                         \\
            160     &  245.30 - 1168.16       & 192                     &   $200.7 \pm 0.7$   &     $234.3 \pm 0.3$     &   $1.084 \pm 0.004$   &   14                                         \\
            200     &  244.30 - 1168.16       & 203                     &   $213.9 \pm 0.7$   &     $233.34 \pm 0.18$   &   $1.063 \pm 0.003$   &   15                                         \\
            250     &  244.30 - 1168.16       & 193                     &   $222.2 \pm 0.7$   &     $234.79 \pm 0.19$   &   $1.026 \pm 0.003$   &   16                                         \\
            297.5   &  244.30 - 1168.16       & 192                     &   $231.3 \pm 0.8$   &     $235.6 \pm 0.2$     &   $1.006 \pm 0.003$   &   17                                         \\
            \hline
    \end{tabular}
    \label{tab:FitSA}
\end{table*}

\section*{Supplementary Material}
Figure S1 which shows $D\eta/T$ vs. $P$ along several low temperature isotherms is available in the Supplementary Material.

\begin{acknowledgments}
We acknowledge funding by the European Research Council under the European Community’s FP7 Grant Agreement 240113; the Institute of Multiscale Science and Technology (Labex iMUST) supported by the French Agence Nationale de la Recherche; and Agence Nationale de la Recherche, Grant No. ANR-19-CE30-0035-01.
\end{acknowledgments}

\section*{Conflict of Interest Statement}
The authors have no conflicts to disclose.

\section*{Author Contribution Statement}
Romain Berthelard: Investigation (equal); Writing - review and editing (equal). Fr\'{e}d\'{e}ric Caupin: Conceptualization (equal); Writing – original draft (equal); Funding Acquisition (lead). Bruno Issenmann: Conceptualization (equal); Funding Acquisition (supporting); Formal analysis (lead); Writing – original draft (equal); Visualization (lead). Alexandre Mussa: Investigation (equal); Writing - review and editing (equal).

\section*{Data Availability Statement}
The data that support the findings of this study are available within the article.

\nocite{*}


\begin{thebibliography}{45}%
\makeatletter
\providecommand \@ifxundefined [1]{%
 \@ifx{#1\undefined}
}%
\providecommand \@ifnum [1]{%
 \ifnum #1\expandafter \@firstoftwo
 \else \expandafter \@secondoftwo
 \fi
}%
\providecommand \@ifx [1]{%
 \ifx #1\expandafter \@firstoftwo
 \else \expandafter \@secondoftwo
 \fi
}%
\providecommand \natexlab [1]{#1}%
\providecommand \enquote  [1]{``#1''}%
\providecommand \bibnamefont  [1]{#1}%
\providecommand \bibfnamefont [1]{#1}%
\providecommand \citenamefont [1]{#1}%
\providecommand \href@noop [0]{\@secondoftwo}%
\providecommand \href [0]{\begingroup \@sanitize@url \@href}%
\providecommand \@href[1]{\@@startlink{#1}\@@href}%
\providecommand \@@href[1]{\endgroup#1\@@endlink}%
\providecommand \@sanitize@url [0]{\catcode `\\12\catcode `\$12\catcode
  `\&12\catcode `\#12\catcode `\^12\catcode `\_12\catcode `\%12\relax}%
\providecommand \@@startlink[1]{}%
\providecommand \@@endlink[0]{}%
\providecommand \url  [0]{\begingroup\@sanitize@url \@url }%
\providecommand \@url [1]{\endgroup\@href {#1}{\urlprefix }}%
\providecommand \urlprefix  [0]{URL }%
\providecommand \Eprint [0]{\href }%
\providecommand \doibase [0]{http://dx.doi.org/}%
\providecommand \selectlanguage [0]{\@gobble}%
\providecommand \bibinfo  [0]{\@secondoftwo}%
\providecommand \bibfield  [0]{\@secondoftwo}%
\providecommand \translation [1]{[#1]}%
\providecommand \BibitemOpen [0]{}%
\providecommand \bibitemStop [0]{}%
\providecommand \bibitemNoStop [0]{.\EOS\space}%
\providecommand \EOS [0]{\spacefactor3000\relax}%
\providecommand \BibitemShut  [1]{\csname bibitem#1\endcsname}%
\let\auto@bib@innerbib\@empty
\bibitem [{\citenamefont {Bett}\ and\ \citenamefont {Cappi}(1965)}]{Bett1966}%
  \BibitemOpen
  \bibfield  {author} {\bibinfo {author} {\bibfnamefont {K.}~\bibnamefont
  {Bett}}\ and\ \bibinfo {author} {\bibfnamefont {J.}~\bibnamefont {Cappi}},\
  }\href@noop {} {\bibfield  {journal} {\bibinfo  {journal} {Nature}\ }\textbf
  {\bibinfo {volume} {207}},\ \bibinfo {pages} {620} (\bibinfo {year}
  {1965})}\BibitemShut {NoStop}%
\bibitem [{\citenamefont {Singh}, \citenamefont {Issenmann},\ and\
  \citenamefont {Caupin}(2017)}]{Singh2017}%
  \BibitemOpen
  \bibfield  {author} {\bibinfo {author} {\bibfnamefont {L.~P.}\ \bibnamefont
  {Singh}}, \bibinfo {author} {\bibfnamefont {B.}~\bibnamefont {Issenmann}}, \
  and\ \bibinfo {author} {\bibfnamefont {F.}~\bibnamefont {Caupin}},\
  }\href@noop {} {\bibfield  {journal} {\bibinfo  {journal} {Proceedings of the
  National Academy of Sciences of the United States of America}\ }\textbf
  {\bibinfo {volume} {114}},\ \bibinfo {pages} {4312} (\bibinfo {year}
  {2017})}\BibitemShut {NoStop}%
\bibitem [{\citenamefont {Chang}\ and\ \citenamefont
  {Sillescu}(1997)}]{Chang1997}%
  \BibitemOpen
  \bibfield  {author} {\bibinfo {author} {\bibfnamefont {I.}~\bibnamefont
  {Chang}}\ and\ \bibinfo {author} {\bibfnamefont {H.}~\bibnamefont
  {Sillescu}},\ }\href@noop {} {\bibfield  {journal} {\bibinfo  {journal} {J.
  Phys. Chem. B}\ }\textbf {\bibinfo {volume} {101}},\ \bibinfo {pages} {8794}
  (\bibinfo {year} {1997})}\BibitemShut {NoStop}%
\bibitem [{\citenamefont {Dehaoui}, \citenamefont {Issenmann},\ and\
  \citenamefont {Caupin}(2015)}]{Dehaoui2015}%
  \BibitemOpen
  \bibfield  {author} {\bibinfo {author} {\bibfnamefont {A.}~\bibnamefont
  {Dehaoui}}, \bibinfo {author} {\bibfnamefont {B.}~\bibnamefont {Issenmann}},
  \ and\ \bibinfo {author} {\bibfnamefont {F.}~\bibnamefont {Caupin}},\
  }\href@noop {} {\bibfield  {journal} {\bibinfo  {journal} {Proceedings of the
  National Academy of Sciences of the United States of America}\ }\textbf
  {\bibinfo {volume} {112}},\ \bibinfo {pages} {12020} (\bibinfo {year}
  {2015})}\BibitemShut {NoStop}%
\bibitem [{\citenamefont {Ragueneau}, \citenamefont {Caupin},\ and\
  \citenamefont {Issenmann}(2022)}]{Ragueneau2022}%
  \BibitemOpen
  \bibfield  {author} {\bibinfo {author} {\bibfnamefont {P.}~\bibnamefont
  {Ragueneau}}, \bibinfo {author} {\bibfnamefont {F.}~\bibnamefont {Caupin}}, \
  and\ \bibinfo {author} {\bibfnamefont {B.}~\bibnamefont {Issenmann}},\
  }\href@noop {} {\bibfield  {journal} {\bibinfo  {journal} {Physical Review
  E}\ }\textbf {\bibinfo {volume} {106}},\ \bibinfo {pages} {014616} (\bibinfo
  {year} {2022})}\BibitemShut {NoStop}%
\bibitem [{\citenamefont {Dubey}\ \emph {et~al.}(2019)\citenamefont {Dubey},
  \citenamefont {Erimban}, \citenamefont {Indra},\ and\ \citenamefont
  {Daschakraborty}}]{Dubey2019}%
  \BibitemOpen
  \bibfield  {author} {\bibinfo {author} {\bibfnamefont {V.}~\bibnamefont
  {Dubey}}, \bibinfo {author} {\bibfnamefont {S.}~\bibnamefont {Erimban}},
  \bibinfo {author} {\bibfnamefont {S.}~\bibnamefont {Indra}}, \ and\ \bibinfo
  {author} {\bibfnamefont {S.}~\bibnamefont {Daschakraborty}},\ }\href@noop {}
  {\bibfield  {journal} {\bibinfo  {journal} {J. Phys. Chem. B}\ }\textbf
  {\bibinfo {volume} {123}},\ \bibinfo {pages} {10089} (\bibinfo {year}
  {2019})}\BibitemShut {NoStop}%
\bibitem [{\citenamefont {Gallo}\ \emph {et~al.}(2016)\citenamefont {Gallo},
  \citenamefont {Amann-Winkel}, \citenamefont {Angell}, \citenamefont
  {Anisimov}, \citenamefont {Caupin}, \citenamefont {Chakravarty},
  \citenamefont {Lascaris}, \citenamefont {Loerting}, \citenamefont
  {Panagiotopoulos}, \citenamefont {Russo}, \citenamefont {Sellberg},
  \citenamefont {Stanley}, \citenamefont {Tanaka}, \citenamefont {Vega},
  \citenamefont {Xu},\ and\ \citenamefont {Pettersson}}]{Gallo2016}%
  \BibitemOpen
  \bibfield  {author} {\bibinfo {author} {\bibfnamefont {P.}~\bibnamefont
  {Gallo}}, \bibinfo {author} {\bibfnamefont {K.}~\bibnamefont {Amann-Winkel}},
  \bibinfo {author} {\bibfnamefont {C.~A.}\ \bibnamefont {Angell}}, \bibinfo
  {author} {\bibfnamefont {M.~A.}\ \bibnamefont {Anisimov}}, \bibinfo {author}
  {\bibfnamefont {F.}~\bibnamefont {Caupin}}, \bibinfo {author} {\bibfnamefont
  {D.}~\bibnamefont {Chakravarty}}, \bibinfo {author} {\bibfnamefont
  {E.}~\bibnamefont {Lascaris}}, \bibinfo {author} {\bibfnamefont
  {T.}~\bibnamefont {Loerting}}, \bibinfo {author} {\bibfnamefont {A.~Z.}\
  \bibnamefont {Panagiotopoulos}}, \bibinfo {author} {\bibfnamefont
  {J.}~\bibnamefont {Russo}}, \bibinfo {author} {\bibfnamefont {J.~A.}\
  \bibnamefont {Sellberg}}, \bibinfo {author} {\bibfnamefont {H.~E.}\
  \bibnamefont {Stanley}}, \bibinfo {author} {\bibfnamefont {H.}~\bibnamefont
  {Tanaka}}, \bibinfo {author} {\bibfnamefont {C.}~\bibnamefont {Vega}},
  \bibinfo {author} {\bibfnamefont {L.}~\bibnamefont {Xu}}, \ and\ \bibinfo
  {author} {\bibfnamefont {L.~G.~M.}\ \bibnamefont {Pettersson}},\ }\href@noop
  {} {\bibfield  {journal} {\bibinfo  {journal} {Chemical Reviews}\ }\textbf
  {\bibinfo {volume} {116}},\ \bibinfo {pages} {7463–7500} (\bibinfo {year}
  {2016})}\BibitemShut {NoStop}%
\bibitem [{\citenamefont {Kumar}\ \emph {et~al.}(2007)\citenamefont {Kumar},
  \citenamefont {Buldyrev}, \citenamefont {Becker}, \citenamefont {Poole},
  \citenamefont {Starr},\ and\ \citenamefont {Stanley}}]{Kumar2007}%
  \BibitemOpen
  \bibfield  {author} {\bibinfo {author} {\bibfnamefont {P.}~\bibnamefont
  {Kumar}}, \bibinfo {author} {\bibfnamefont {S.~V.}\ \bibnamefont {Buldyrev}},
  \bibinfo {author} {\bibfnamefont {S.~R.}\ \bibnamefont {Becker}}, \bibinfo
  {author} {\bibfnamefont {P.~H.}\ \bibnamefont {Poole}}, \bibinfo {author}
  {\bibfnamefont {F.~W.}\ \bibnamefont {Starr}}, \ and\ \bibinfo {author}
  {\bibfnamefont {H.~E.}\ \bibnamefont {Stanley}},\ }\href@noop {} {\bibfield
  {journal} {\bibinfo  {journal} {Proceedings of the National Academy of
  Sciences USA}\ }\textbf {\bibinfo {volume} {104}},\ \bibinfo {pages} {9575}
  (\bibinfo {year} {2007})}\BibitemShut {NoStop}%
\bibitem [{\citenamefont {Montero~de Hijes}\ \emph {et~al.}(2018)\citenamefont
  {Montero~de Hijes}, \citenamefont {Sanz}, \citenamefont {Joly}, \citenamefont
  {Valeriani},\ and\ \citenamefont {Caupin}}]{Montero2018}%
  \BibitemOpen
  \bibfield  {author} {\bibinfo {author} {\bibfnamefont {P.}~\bibnamefont
  {Montero~de Hijes}}, \bibinfo {author} {\bibfnamefont {E.}~\bibnamefont
  {Sanz}}, \bibinfo {author} {\bibfnamefont {L.}~\bibnamefont {Joly}}, \bibinfo
  {author} {\bibfnamefont {C.}~\bibnamefont {Valeriani}}, \ and\ \bibinfo
  {author} {\bibfnamefont {F.}~\bibnamefont {Caupin}},\ }\href@noop {}
  {\bibfield  {journal} {\bibinfo  {journal} {J. Chem. Phys.}\ }\textbf
  {\bibinfo {volume} {149}},\ \bibinfo {pages} {094503} (\bibinfo {year}
  {2018})}\BibitemShut {NoStop}%
\bibitem [{\citenamefont {Hallet}(1963)}]{Hallet1963}%
  \BibitemOpen
  \bibfield  {author} {\bibinfo {author} {\bibfnamefont {J.}~\bibnamefont
  {Hallet}},\ }\href@noop {} {\bibfield  {journal} {\bibinfo  {journal} {Proc.
  Phys. Soc.}\ }\textbf {\bibinfo {volume} {82}},\ \bibinfo {pages} {1046}
  (\bibinfo {year} {1963})}\BibitemShut {NoStop}%
\bibitem [{\citenamefont {Yu.A.Osipov}, \citenamefont {B.V.Zhelezn'yi},\ and\
  \citenamefont {.Bondarenko}(1977)}]{Osipov1977}%
  \BibitemOpen
  \bibfield  {author} {\bibinfo {author} {\bibnamefont {Yu.A.Osipov}}, \bibinfo
  {author} {\bibnamefont {B.V.Zhelezn'yi}}, \ and\ \bibinfo {author}
  {\bibfnamefont {N.}~\bibnamefont {.Bondarenko}},\ }\href@noop {} {\bibfield
  {journal} {\bibinfo  {journal} {Russian Journal of Physical Chemistry}\
  }\textbf {\bibinfo {volume} {51}},\ \bibinfo {pages} {748} (\bibinfo {year}
  {1977})}\BibitemShut {NoStop}%
\bibitem [{\citenamefont {Cho}\ \emph {et~al.}(1999)\citenamefont {Cho},
  \citenamefont {Urquidi}, \citenamefont {Singh},\ and\ \citenamefont
  {Robinson}}]{Cho1999}%
  \BibitemOpen
  \bibfield  {author} {\bibinfo {author} {\bibfnamefont {C.~H.}\ \bibnamefont
  {Cho}}, \bibinfo {author} {\bibfnamefont {J.}~\bibnamefont {Urquidi}},
  \bibinfo {author} {\bibfnamefont {S.}~\bibnamefont {Singh}}, \ and\ \bibinfo
  {author} {\bibfnamefont {G.~W.}\ \bibnamefont {Robinson}},\ }\href@noop {}
  {\bibfield  {journal} {\bibinfo  {journal} {J. Phys. Chem. B}\ }\textbf
  {\bibinfo {volume} {103}},\ \bibinfo {pages} {1991} (\bibinfo {year}
  {1999})}\BibitemShut {NoStop}%
\bibitem [{\citenamefont {Cerbino}\ and\ \citenamefont
  {Trappe}(2008)}]{Cerbino2008}%
  \BibitemOpen
  \bibfield  {author} {\bibinfo {author} {\bibfnamefont {R.}~\bibnamefont
  {Cerbino}}\ and\ \bibinfo {author} {\bibfnamefont {V.}~\bibnamefont
  {Trappe}},\ }\href@noop {} {\bibfield  {journal} {\bibinfo  {journal}
  {Physical Review Letters}\ }\textbf {\bibinfo {volume} {100}},\ \bibinfo
  {pages} {188102} (\bibinfo {year} {2008})}\BibitemShut {NoStop}%
\bibitem [{\citenamefont {Giavazzi}\ \emph {et~al.}(2009)\citenamefont
  {Giavazzi}, \citenamefont {Brogioli}, \citenamefont {Trappe}, \citenamefont
  {Bellini},\ and\ \citenamefont {Cerbino}}]{Giavazzi2009}%
  \BibitemOpen
  \bibfield  {author} {\bibinfo {author} {\bibfnamefont {F.}~\bibnamefont
  {Giavazzi}}, \bibinfo {author} {\bibfnamefont {D.}~\bibnamefont {Brogioli}},
  \bibinfo {author} {\bibfnamefont {V.}~\bibnamefont {Trappe}}, \bibinfo
  {author} {\bibfnamefont {T.}~\bibnamefont {Bellini}}, \ and\ \bibinfo
  {author} {\bibfnamefont {R.}~\bibnamefont {Cerbino}},\ }\href@noop {}
  {\bibfield  {journal} {\bibinfo  {journal} {Phys. Rev. E}\ }\textbf {\bibinfo
  {volume} {80}},\ \bibinfo {pages} {031403} (\bibinfo {year}
  {2009})}\BibitemShut {NoStop}%
\bibitem [{\citenamefont {Patel}\ \emph {et~al.}(2004)\citenamefont {Patel},
  \citenamefont {Jerkovich}, \citenamefont {Link},\ and\ \citenamefont
  {Jorgenson}}]{Patel2004}%
  \BibitemOpen
  \bibfield  {author} {\bibinfo {author} {\bibfnamefont {K.~D.}\ \bibnamefont
  {Patel}}, \bibinfo {author} {\bibfnamefont {A.~D.}\ \bibnamefont
  {Jerkovich}}, \bibinfo {author} {\bibfnamefont {J.~C.}\ \bibnamefont {Link}},
  \ and\ \bibinfo {author} {\bibfnamefont {J.~W.}\ \bibnamefont {Jorgenson}},\
  }\href@noop {} {\bibfield  {journal} {\bibinfo  {journal} {Analytical
  Chemistry}\ }\textbf {\bibinfo {volume} {76}},\ \bibinfo {pages} {5777}
  (\bibinfo {year} {2004})}\BibitemShut {NoStop}%
\bibitem [{\citenamefont {Woolf}(1974)}]{Woolf1974}%
  \BibitemOpen
  \bibfield  {author} {\bibinfo {author} {\bibfnamefont {L.~A.}\ \bibnamefont
  {Woolf}},\ }\href@noop {} {\bibfield  {journal} {\bibinfo  {journal} {The
  Journal of Chemical Physics}\ }\textbf {\bibinfo {volume} {61}},\ \bibinfo
  {pages} {1600} (\bibinfo {year} {1974})}\BibitemShut {NoStop}%
\bibitem [{\citenamefont {Woolf}(1975)}]{Woolf1975}%
  \BibitemOpen
  \bibfield  {author} {\bibinfo {author} {\bibfnamefont {L.~A.}\ \bibnamefont
  {Woolf}},\ }\href@noop {} {\bibfield  {journal} {\bibinfo  {journal} {Journal
  of the Chemical Society, Faraday Transactions 1}\ }\textbf {\bibinfo {volume}
  {71}},\ \bibinfo {pages} {784} (\bibinfo {year} {1975})}\BibitemShut
  {NoStop}%
\bibitem [{\citenamefont {Angell}\ \emph {et~al.}(1976)\citenamefont {Angell},
  \citenamefont {Finch}, \citenamefont {Woolf},\ and\ \citenamefont
  {Bach}}]{Angell1976}%
  \BibitemOpen
  \bibfield  {author} {\bibinfo {author} {\bibfnamefont {C.~A.}\ \bibnamefont
  {Angell}}, \bibinfo {author} {\bibfnamefont {E.~D.}\ \bibnamefont {Finch}},
  \bibinfo {author} {\bibfnamefont {L.~A.}\ \bibnamefont {Woolf}}, \ and\
  \bibinfo {author} {\bibfnamefont {P.}~\bibnamefont {Bach}},\ }\href@noop {}
  {\bibfield  {journal} {\bibinfo  {journal} {The Journal of Chemical Physics}\
  }\textbf {\bibinfo {volume} {65}},\ \bibinfo {pages} {3063} (\bibinfo {year}
  {1976})}\BibitemShut {NoStop}%
\bibitem [{\citenamefont {Krynicki}, \citenamefont {Green},\ and\ \citenamefont
  {Sawyer}(1978)}]{Krynicki1978}%
  \BibitemOpen
  \bibfield  {author} {\bibinfo {author} {\bibfnamefont {K.}~\bibnamefont
  {Krynicki}}, \bibinfo {author} {\bibfnamefont {C.~D.}\ \bibnamefont {Green}},
  \ and\ \bibinfo {author} {\bibfnamefont {D.~W.}\ \bibnamefont {Sawyer}},\
  }\href@noop {} {\bibfield  {journal} {\bibinfo  {journal} {Faraday
  Discussions of the Chemical Society}\ }\textbf {\bibinfo {volume} {66}},\
  \bibinfo {pages} {199} (\bibinfo {year} {1978})}\BibitemShut {NoStop}%
\bibitem [{\citenamefont {Harris}\ and\ \citenamefont
  {Woolf}(1980)}]{Harris1980}%
  \BibitemOpen
  \bibfield  {author} {\bibinfo {author} {\bibfnamefont {K.~B.}\ \bibnamefont
  {Harris}}\ and\ \bibinfo {author} {\bibfnamefont {L.~A.}\ \bibnamefont
  {Woolf}},\ }\href@noop {} {\bibfield  {journal} {\bibinfo  {journal} {Journal
  of the Chemical Society, Faraday Transactions 1}\ }\textbf {\bibinfo {volume}
  {76}},\ \bibinfo {pages} {377} (\bibinfo {year} {1980})}\BibitemShut
  {NoStop}%
\bibitem [{\citenamefont {Easteal}, \citenamefont {Edge},\ and\ \citenamefont
  {Woolf}(1984)}]{Easteal1984}%
  \BibitemOpen
  \bibfield  {author} {\bibinfo {author} {\bibfnamefont {A.~J.}\ \bibnamefont
  {Easteal}}, \bibinfo {author} {\bibfnamefont {A.~V.~J.}\ \bibnamefont
  {Edge}}, \ and\ \bibinfo {author} {\bibfnamefont {L.~A.}\ \bibnamefont
  {Woolf}},\ }\href@noop {} {\bibfield  {journal} {\bibinfo  {journal} {The
  Journal of Physical Chemistry}\ }\textbf {\bibinfo {volume} {88}},\ \bibinfo
  {pages} {6060} (\bibinfo {year} {1984})}\BibitemShut {NoStop}%
\bibitem [{\citenamefont {Baker}\ and\ \citenamefont
  {Jonas}(1985)}]{Baker1985}%
  \BibitemOpen
  \bibfield  {author} {\bibinfo {author} {\bibfnamefont {E.~S.}\ \bibnamefont
  {Baker}}\ and\ \bibinfo {author} {\bibfnamefont {J.}~\bibnamefont {Jonas}},\
  }\href@noop {} {\bibfield  {journal} {\bibinfo  {journal} {The Journal of
  Physical Chemistry}\ }\textbf {\bibinfo {volume} {89}},\ \bibinfo {pages}
  {1730} (\bibinfo {year} {1985})}\BibitemShut {NoStop}%
\bibitem [{\citenamefont {Prielmeier}\ \emph {et~al.}(1988)\citenamefont
  {Prielmeier}, \citenamefont {Lang}, \citenamefont {Speedy},\ and\
  \citenamefont {L\"udemann}}]{Prielmeier1988}%
  \BibitemOpen
  \bibfield  {author} {\bibinfo {author} {\bibfnamefont {F.~X.}\ \bibnamefont
  {Prielmeier}}, \bibinfo {author} {\bibfnamefont {E.~W.}\ \bibnamefont
  {Lang}}, \bibinfo {author} {\bibfnamefont {R.~J.}\ \bibnamefont {Speedy}}, \
  and\ \bibinfo {author} {\bibfnamefont {H.-D.}\ \bibnamefont {L\"udemann}},\
  }\href@noop {} {\bibfield  {journal} {\bibinfo  {journal} {Ber. Bunsenges.
  Phys. Chem.}\ }\textbf {\bibinfo {volume} {92}},\ \bibinfo {pages} {1111}
  (\bibinfo {year} {1988})}\BibitemShut {NoStop}%
\bibitem [{\citenamefont {Harris}\ and\ \citenamefont
  {Newitt}(1997)}]{Harris1997}%
  \BibitemOpen
  \bibfield  {author} {\bibinfo {author} {\bibfnamefont {K.~R.}\ \bibnamefont
  {Harris}}\ and\ \bibinfo {author} {\bibfnamefont {P.~J.}\ \bibnamefont
  {Newitt}},\ }\href@noop {} {\bibfield  {journal} {\bibinfo  {journal}
  {Journal of Chemical and Engineering Data}\ }\textbf {\bibinfo {volume}
  {42}},\ \bibinfo {pages} {346} (\bibinfo {year} {1997})}\BibitemShut
  {NoStop}%
\bibitem [{\citenamefont {Huber}\ \emph {et~al.}(2009)\citenamefont {Huber},
  \citenamefont {Perkins}, \citenamefont {Laesecke}, \citenamefont {Friend},
  \citenamefont {Sengers}, \citenamefont {Assael}, \citenamefont {Metaxa},
  \citenamefont {Vogel}, \citenamefont {Mares},\ and\ \citenamefont
  {Miyagawa}}]{Huber2009}%
  \BibitemOpen
  \bibfield  {author} {\bibinfo {author} {\bibfnamefont {M.~L.}\ \bibnamefont
  {Huber}}, \bibinfo {author} {\bibfnamefont {R.~A.}\ \bibnamefont {Perkins}},
  \bibinfo {author} {\bibfnamefont {A.}~\bibnamefont {Laesecke}}, \bibinfo
  {author} {\bibfnamefont {D.~G.}\ \bibnamefont {Friend}}, \bibinfo {author}
  {\bibfnamefont {J.~V.}\ \bibnamefont {Sengers}}, \bibinfo {author}
  {\bibfnamefont {M.~J.}\ \bibnamefont {Assael}}, \bibinfo {author}
  {\bibfnamefont {I.~N.}\ \bibnamefont {Metaxa}}, \bibinfo {author}
  {\bibfnamefont {E.}~\bibnamefont {Vogel}}, \bibinfo {author} {\bibfnamefont
  {R.}~\bibnamefont {Mares}}, \ and\ \bibinfo {author} {\bibfnamefont
  {K.}~\bibnamefont {Miyagawa}},\ }\href@noop {} {\bibfield  {journal}
  {\bibinfo  {journal} {J. Phys. Chem. Ref. Data}\ }\textbf {\bibinfo {volume}
  {38}},\ \bibinfo {pages} {101} (\bibinfo {year} {2009})}\BibitemShut
  {NoStop}%
\bibitem [{\citenamefont {Gillen}, \citenamefont {Douglass},\ and\
  \citenamefont {Hoch}(1972)}]{Gillen1972}%
  \BibitemOpen
  \bibfield  {author} {\bibinfo {author} {\bibfnamefont {K.~T.}\ \bibnamefont
  {Gillen}}, \bibinfo {author} {\bibfnamefont {D.~C.}\ \bibnamefont
  {Douglass}}, \ and\ \bibinfo {author} {\bibfnamefont {M.~J.~R.}\ \bibnamefont
  {Hoch}},\ }\href@noop {} {\bibfield  {journal} {\bibinfo  {journal} {The
  Journal of Chemical Physics}\ }\textbf {\bibinfo {volume} {57}},\ \bibinfo
  {pages} {5117} (\bibinfo {year} {1972})}\BibitemShut {NoStop}%
\bibitem [{\citenamefont {Price}, \citenamefont {Ide},\ and\ \citenamefont
  {Arata}(1999)}]{Price1999}%
  \BibitemOpen
  \bibfield  {author} {\bibinfo {author} {\bibfnamefont {W.~S.}\ \bibnamefont
  {Price}}, \bibinfo {author} {\bibfnamefont {H.}~\bibnamefont {Ide}}, \ and\
  \bibinfo {author} {\bibfnamefont {Y.}~\bibnamefont {Arata}},\ }\href@noop {}
  {\bibfield  {journal} {\bibinfo  {journal} {Journal of Physical Chemistry A}\
  }\textbf {\bibinfo {volume} {103}},\ \bibinfo {pages} {448} (\bibinfo {year}
  {1999})}\BibitemShut {NoStop}%
\bibitem [{\citenamefont {Su\'arez-Iglesias}\ \emph {et~al.}(2015)\citenamefont
  {Su\'arez-Iglesias}, \citenamefont {Medina}, \citenamefont {de~los
  \'Angeles~Sanz}, \citenamefont {Pizarro},\ and\ \citenamefont
  {Bueno}}]{Suarez2015}%
  \BibitemOpen
  \bibfield  {author} {\bibinfo {author} {\bibfnamefont {O.}~\bibnamefont
  {Su\'arez-Iglesias}}, \bibinfo {author} {\bibfnamefont {I.}~\bibnamefont
  {Medina}}, \bibinfo {author} {\bibfnamefont {M.}~\bibnamefont {de~los
  \'Angeles~Sanz}}, \bibinfo {author} {\bibfnamefont {C.}~\bibnamefont
  {Pizarro}}, \ and\ \bibinfo {author} {\bibfnamefont {J.~L.}\ \bibnamefont
  {Bueno}},\ }\href@noop {} {\bibfield  {journal} {\bibinfo  {journal} {J.
  Chem. Eng. Data}\ }\textbf {\bibinfo {volume} {60}},\ \bibinfo {pages} {2757}
  (\bibinfo {year} {2015})}\BibitemShut {NoStop}%
\bibitem [{\citenamefont {Mills}(1973)}]{Mills1973}%
  \BibitemOpen
  \bibfield  {author} {\bibinfo {author} {\bibfnamefont {R.}~\bibnamefont
  {Mills}},\ }\href@noop {} {\bibfield  {journal} {\bibinfo  {journal} {The
  Journal of Physical Chemistry}\ }\textbf {\bibinfo {volume} {77}},\ \bibinfo
  {pages} {685} (\bibinfo {year} {1973})}\BibitemShut {NoStop}%
\bibitem [{\citenamefont {for the Properties~of Water}\ and\ \citenamefont
  {Steam}(2011)}]{IAPWS2011}%
  \BibitemOpen
  \bibfield  {author} {\bibinfo {author} {\bibfnamefont {T.~I.~A.}\
  \bibnamefont {for the Properties~of Water}}\ and\ \bibinfo {author}
  {\bibnamefont {Steam}},\ }\href@noop {} {\enquote {\bibinfo {title}
  {{R}evised {R}elease on the {P}ressure along the {M}elting and {S}ublimation
  {C}urves of {O}rdinary {W}ater {S}ubstance},}\ }\bibinfo {type} {Tech. Rep.}\
  \bibinfo {number} {{R}14-08}\ (\bibinfo  {institution} {International
  Association for the Properties of Water and Steam},\ \bibinfo {year}
  {2011})\BibitemShut {NoStop}%
\bibitem [{\citenamefont {for the Properties~of Water}\ and\ \citenamefont
  {Steam}(2015)}]{IAPWS2015}%
  \BibitemOpen
  \bibfield  {author} {\bibinfo {author} {\bibfnamefont {T.~I.~A.}\
  \bibnamefont {for the Properties~of Water}}\ and\ \bibinfo {author}
  {\bibnamefont {Steam}},\ }\href@noop {} {\enquote {\bibinfo {title}
  {{G}uideline on {T}hermodynamic {P}roperties of {S}upercooled {W}ater},}\
  }\bibinfo {type} {Tech. Rep.}\ \bibinfo {number} {{G}12-15}\ (\bibinfo
  {institution} {International Association for the Properties of Water and
  Steam},\ \bibinfo {year} {2015})\BibitemShut {NoStop}%
\bibitem [{\citenamefont {Caupin}\ and\ \citenamefont
  {Anisimov}(2019)}]{Caupin2019}%
  \BibitemOpen
  \bibfield  {author} {\bibinfo {author} {\bibfnamefont {F.}~\bibnamefont
  {Caupin}}\ and\ \bibinfo {author} {\bibfnamefont {M.~A.}\ \bibnamefont
  {Anisimov}},\ }\href@noop {} {\bibfield  {journal} {\bibinfo  {journal} {J.
  Chem. Phys.}\ }\textbf {\bibinfo {volume} {151}},\ \bibinfo {pages} {034503}
  (\bibinfo {year} {2019})}\BibitemShut {NoStop}%
\bibitem [{\citenamefont {R{\"o}ntgen}(1884)}]{Roentgen1884}%
  \BibitemOpen
  \bibfield  {author} {\bibinfo {author} {\bibfnamefont {W.~C.}\ \bibnamefont
  {R{\"o}ntgen}},\ }\href {\doibase 10.1002/andp.18842580804} {\bibfield
  {journal} {\bibinfo  {journal} {Ann. Phys.}\ }\textbf {\bibinfo {volume}
  {258}},\ \bibinfo {pages} {510} (\bibinfo {year} {1884})}\BibitemShut
  {NoStop}%
\bibitem [{\citenamefont {Warburg}\ and\ \citenamefont
  {Sachs}(1884)}]{Warburg1884}%
  \BibitemOpen
  \bibfield  {author} {\bibinfo {author} {\bibfnamefont {E.}~\bibnamefont
  {Warburg}}\ and\ \bibinfo {author} {\bibfnamefont {J.}~\bibnamefont
  {Sachs}},\ }\href {\doibase 10.1002/andp.18842580805} {\bibfield  {journal}
  {\bibinfo  {journal} {Ann. Phys.}\ }\textbf {\bibinfo {volume} {258}},\
  \bibinfo {pages} {518} (\bibinfo {year} {1884})}\BibitemShut {NoStop}%
\bibitem [{\citenamefont {Amann-Winkel}\ \emph {et~al.}(2013)\citenamefont
  {Amann-Winkel}, \citenamefont {Gainaru}, \citenamefont {Handle},
  \citenamefont {Seidl}, \citenamefont {Nelson}, \citenamefont {Böhmer},\ and\
  \citenamefont {Loerting}}]{AmannWinkel2013}%
  \BibitemOpen
  \bibfield  {author} {\bibinfo {author} {\bibfnamefont {K.}~\bibnamefont
  {Amann-Winkel}}, \bibinfo {author} {\bibfnamefont {C.}~\bibnamefont
  {Gainaru}}, \bibinfo {author} {\bibfnamefont {P.~H.}\ \bibnamefont {Handle}},
  \bibinfo {author} {\bibfnamefont {M.}~\bibnamefont {Seidl}}, \bibinfo
  {author} {\bibfnamefont {H.}~\bibnamefont {Nelson}}, \bibinfo {author}
  {\bibfnamefont {R.}~\bibnamefont {Böhmer}}, \ and\ \bibinfo {author}
  {\bibfnamefont {T.}~\bibnamefont {Loerting}},\ }\href@noop {} {\bibfield
  {journal} {\bibinfo  {journal} {PNAS}\ }\textbf {\bibinfo {volume} {110}},\
  \bibinfo {pages} {17720} (\bibinfo {year} {2013})}\BibitemShut {NoStop}%
\bibitem [{\citenamefont {Vega}\ and\ \citenamefont
  {Abascal}(2011)}]{Vega2011}%
  \BibitemOpen
  \bibfield  {author} {\bibinfo {author} {\bibfnamefont {C.}~\bibnamefont
  {Vega}}\ and\ \bibinfo {author} {\bibfnamefont {J.~L.~F.}\ \bibnamefont
  {Abascal}},\ }\href {\doibase 10.1039/c1cp22168j} {\bibfield  {journal}
  {\bibinfo  {journal} {Phys. Chem. Chem. Phys.}\ }\textbf {\bibinfo {volume}
  {13}},\ \bibinfo {pages} {19663} (\bibinfo {year} {2011})}\BibitemShut
  {NoStop}%
\bibitem [{\citenamefont {Biddle}\ \emph {et~al.}(2017)\citenamefont {Biddle},
  \citenamefont {Singh}, \citenamefont {Sparano}, \citenamefont {Ricci},
  \citenamefont {Gonzalez}, \citenamefont {Valeriani}, \citenamefont {Abascal},
  \citenamefont {Debenedetti}, \citenamefont {Anisimov},\ and\ \citenamefont
  {Caupin}}]{Biddle2017}%
  \BibitemOpen
  \bibfield  {author} {\bibinfo {author} {\bibfnamefont {J.~W.}\ \bibnamefont
  {Biddle}}, \bibinfo {author} {\bibfnamefont {R.~S.}\ \bibnamefont {Singh}},
  \bibinfo {author} {\bibfnamefont {E.~M.}\ \bibnamefont {Sparano}}, \bibinfo
  {author} {\bibfnamefont {F.}~\bibnamefont {Ricci}}, \bibinfo {author}
  {\bibfnamefont {M.~A.}\ \bibnamefont {Gonzalez}}, \bibinfo {author}
  {\bibfnamefont {C.}~\bibnamefont {Valeriani}}, \bibinfo {author}
  {\bibfnamefont {J.~L.~F.}\ \bibnamefont {Abascal}}, \bibinfo {author}
  {\bibfnamefont {P.~G.}\ \bibnamefont {Debenedetti}}, \bibinfo {author}
  {\bibfnamefont {M.~A.}\ \bibnamefont {Anisimov}}, \ and\ \bibinfo {author}
  {\bibfnamefont {F.}~\bibnamefont {Caupin}},\ }\href@noop {} {\bibfield
  {journal} {\bibinfo  {journal} {J. Chem. Phys.}\ }\textbf {\bibinfo {volume}
  {146}},\ \bibinfo {pages} {034502} (\bibinfo {year} {2017})}\BibitemShut
  {NoStop}%
\bibitem [{\citenamefont {Debenedetti}, \citenamefont {Sciortino},\ and\
  \citenamefont {Zerze}(2020)}]{Debenedetti2020}%
  \BibitemOpen
  \bibfield  {author} {\bibinfo {author} {\bibfnamefont {P.~G.}\ \bibnamefont
  {Debenedetti}}, \bibinfo {author} {\bibfnamefont {F.}~\bibnamefont
  {Sciortino}}, \ and\ \bibinfo {author} {\bibfnamefont {G.~H.}\ \bibnamefont
  {Zerze}},\ }\href {\doibase 10.1126/science.abb9796} {\bibfield  {journal}
  {\bibinfo  {journal} {Science}\ }\textbf {\bibinfo {volume} {369}},\ \bibinfo
  {pages} {289} (\bibinfo {year} {2020})}\BibitemShut {NoStop}%
\bibitem [{\citenamefont {Holten}\ and\ \citenamefont
  {Anisimov}(2012)}]{Holten2012}%
  \BibitemOpen
  \bibfield  {author} {\bibinfo {author} {\bibfnamefont {V.}~\bibnamefont
  {Holten}}\ and\ \bibinfo {author} {\bibfnamefont {M.~A.}\ \bibnamefont
  {Anisimov}},\ }\href {\doibase 10.1038/srep00713} {\bibfield  {journal}
  {\bibinfo  {journal} {Sci. Rep.}\ }\textbf {\bibinfo {volume} {2}},\ \bibinfo
  {pages} {713} (\bibinfo {year} {2012})}\BibitemShut {NoStop}%
\bibitem [{\citenamefont {Du{\v s}ka}(2020)}]{Duska2020}%
  \BibitemOpen
  \bibfield  {author} {\bibinfo {author} {\bibfnamefont {M.}~\bibnamefont
  {Du{\v s}ka}},\ }\href {\doibase 10.1063/5.0006431} {\bibfield  {journal}
  {\bibinfo  {journal} {J. Chem. Phys.}\ }\textbf {\bibinfo {volume} {152}},\
  \bibinfo {pages} {174501} (\bibinfo {year} {2020})}\BibitemShut {NoStop}%
\bibitem [{\citenamefont {Mishima}\ and\ \citenamefont
  {Stanley}(1998)}]{Mishima1998}%
  \BibitemOpen
  \bibfield  {author} {\bibinfo {author} {\bibfnamefont {O.}~\bibnamefont
  {Mishima}}\ and\ \bibinfo {author} {\bibfnamefont {H.~E.}\ \bibnamefont
  {Stanley}},\ }\href {\doibase 10.1038/32386} {\bibfield  {journal} {\bibinfo
  {journal} {Nature}\ }\textbf {\bibinfo {volume} {392}},\ \bibinfo {pages}
  {164} (\bibinfo {year} {1998})}\BibitemShut {NoStop}%
\bibitem [{\citenamefont {Mishima}\ and\ \citenamefont
  {Sumita}(2023)}]{Mishima2023}%
  \BibitemOpen
  \bibfield  {author} {\bibinfo {author} {\bibfnamefont {O.}~\bibnamefont
  {Mishima}}\ and\ \bibinfo {author} {\bibfnamefont {T.}~\bibnamefont
  {Sumita}},\ }\href {\doibase 10.1021/acs.jpcb.2c08342} {\bibfield  {journal}
  {\bibinfo  {journal} {J. Phys. Chem. B}\ }\textbf {\bibinfo {volume} {127}},\
  \bibinfo {pages} {1414} (\bibinfo {year} {2023})}\BibitemShut {NoStop}%
\bibitem [{\citenamefont {Shi}\ and\ \citenamefont {Tanaka}(2020)}]{Shi2020}%
  \BibitemOpen
  \bibfield  {author} {\bibinfo {author} {\bibfnamefont {R.}~\bibnamefont
  {Shi}}\ and\ \bibinfo {author} {\bibfnamefont {H.}~\bibnamefont {Tanaka}},\
  }\href {\doibase 10.1073/pnas.2008426117} {\bibfield  {journal} {\bibinfo
  {journal} {Proc. Natl. Acad. Sci.}\ }\textbf {\bibinfo {volume} {117}},\
  \bibinfo {pages} {26591} (\bibinfo {year} {2020})}\BibitemShut {NoStop}%
\bibitem [{\citenamefont {Bachler}, \citenamefont {Giebelmann},\ and\
  \citenamefont {Loerting}(2021)}]{Bachler2021}%
  \BibitemOpen
  \bibfield  {author} {\bibinfo {author} {\bibfnamefont {J.}~\bibnamefont
  {Bachler}}, \bibinfo {author} {\bibfnamefont {J.}~\bibnamefont {Giebelmann}},
  \ and\ \bibinfo {author} {\bibfnamefont {T.}~\bibnamefont {Loerting}},\
  }\href {\doibase 10.1073/pnas.2108194118} {\bibfield  {journal} {\bibinfo
  {journal} {Proc. Natl. Acad. Sci.}\ }\textbf {\bibinfo {volume} {118}},\
  \bibinfo {pages} {e2108194118} (\bibinfo {year} {2021})}\BibitemShut
  {NoStop}%
\bibitem [{\citenamefont {Weingärtner}(1982)}]{Weingartner1982}%
  \BibitemOpen
  \bibfield  {author} {\bibinfo {author} {\bibfnamefont {H.}~\bibnamefont
  {Weingärtner}},\ }\href@noop {} {\bibfield  {journal} {\bibinfo  {journal}
  {Zeitschrift für Physikalische Chemie Neue Folge}\ }\textbf {\bibinfo
  {volume} {132}},\ \bibinfo {pages} {129} (\bibinfo {year}
  {1982})}\BibitemShut {NoStop}%
\end{thebibliography}
\end{document}



\title{Supplementary material for ``Viscosity and Stokes-Einstein relation in deeply supercooled water under pressure''}
\author{Alexandre Mussa}

\author{Romain Berthelard}
\thanks{A.M. and R.B. contributed equally to this work.}%

\author{Fr\'{e}d\'{e}ric Caupin}

\author{Bruno Issenmann}
\email{frederic.caupin@univ-lyon1.fr, bruno.issenmann@univ-lyon1.fr}

\affiliation{%
{Institut Lumi\`ere Mati\`ere, Universit\'e de Lyon, Universit\'e Claude Bernard Lyon 1, CNRS, Institut Universitaire de France, F-69622 Villeurbanne, France
}%
}

\date{18 September 2023}
\begin{abstract}
\end{abstract}
\maketitle

\begin{figure*}[ttt]
    \centering
    \includegraphics[width=0.8\textwidth]{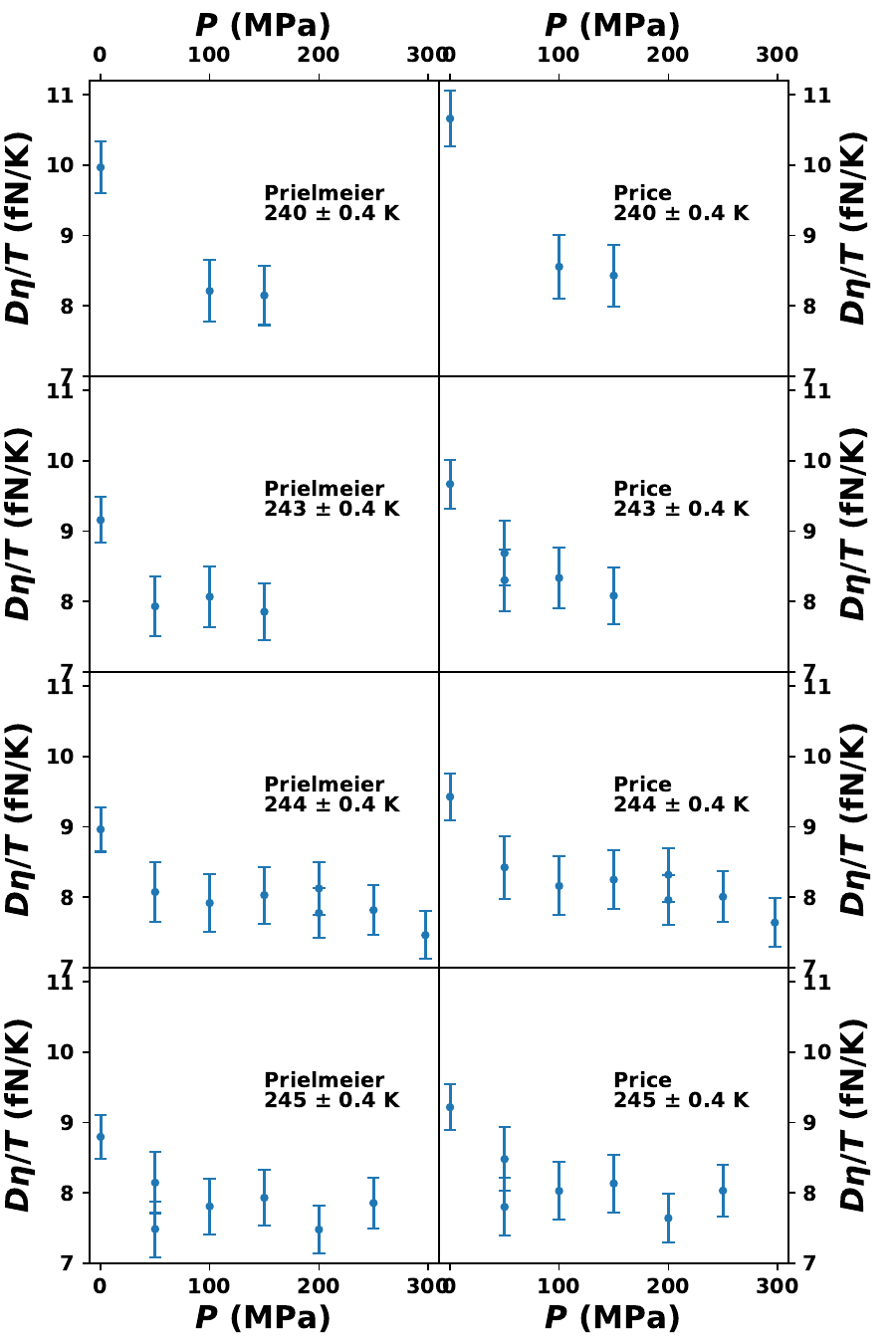}
    \caption{Experimental values for the Stokes-Einstein ratio, $D\eta/T$, as a function of pressure for four different isotherms. Data were considered to belong to the same isotherm when the actual measurement temperature was less than $0.4\,\mathrm{K}$ away from the isotherm temperature. The two columns correspond to two different choices for temperature rescaling (see main text for details).}
    \label{fig:ExperimentalSetup}
\end{figure*}